\pdfoutput=1

\documentclass[12pt,a4paper]{article}

\usepackage{ifthen} 
\newboolean{pdflatex}
\setboolean{pdflatex}{true} 

\newboolean{articletitles}
\setboolean{articletitles}{true} 

\newboolean{uprightparticles}
\setboolean{uprightparticles}{true} 

\newboolean{inbibliography}
\setboolean{inbibliography}{false} 

\usepackage{microtype}
\usepackage{lineno}  
\usepackage{xspace} 

\usepackage{graphicx}  
\usepackage{color}
\usepackage{colortbl}
\graphicspath{{./figs/}} 

\usepackage{amsmath} 
\usepackage{amssymb}
\usepackage{amsfonts}
\usepackage{upgreek} 

\newcommand*\patchAmsMathEnvironmentForLineno[1]{%
\expandafter\let\csname old#1\expandafter\endcsname\csname #1\endcsname
\expandafter\let\csname oldend#1\expandafter\endcsname\csname
end#1\endcsname
 \renewenvironment{#1}%
   {\linenomath\csname old#1\endcsname}%
   {\csname oldend#1\endcsname\endlinenomath}%
}
\newcommand*\patchBothAmsMathEnvironmentsForLineno[1]{%
  \patchAmsMathEnvironmentForLineno{#1}%
  \patchAmsMathEnvironmentForLineno{#1*}%
}
\AtBeginDocument{%
\patchBothAmsMathEnvironmentsForLineno{equation}%
\patchBothAmsMathEnvironmentsForLineno{align}%
\patchBothAmsMathEnvironmentsForLineno{flalign}%
\patchBothAmsMathEnvironmentsForLineno{alignat}%
\patchBothAmsMathEnvironmentsForLineno{gather}%
\patchBothAmsMathEnvironmentsForLineno{multline}%
}

\usepackage{hyperref}    
\usepackage[all]{hypcap} 

\def\lhcb {\mbox{LHCb}\xspace}

\ifthenelse{\boolean{uprightparticles}}%
{

 \def\Pmu         {\ensuremath{\upmu}\xspace}

 \def\Ptau        {\ensuremath{\uptau}\xspace}

 \def\PDelta      {\ensuremath{\Delta}\xspace}                 
 \def\PXi      {\ensuremath{\Xi}\xspace}                 
 \def\PLambda      {\ensuremath{\Lambda}\xspace}                 
 \def\PSigma      {\ensuremath{\Sigma}\xspace}                 
 \def\POmega      {\ensuremath{\Omega}\xspace}                 
 \def\PUpsilon      {\ensuremath{\Upsilon}\xspace}                 
 
 \def\PB      {\ensuremath{\mathrm{B}}\xspace}                 
                  
 \def\PD      {\ensuremath{\mathrm{D}}\xspace}

 \def\PK      {\ensuremath{\mathrm{K}}\xspace}

 \def\PW      {\ensuremath{\mathrm{W}}\xspace}

 \def\PZ      {\ensuremath{\mathrm{Z}}\xspace}                 
                  
 \def\Pb      {\ensuremath{\mathrm{b}}\xspace}                 
 \def\Pc      {\ensuremath{\mathrm{c}}\xspace}

 \def\Pi      {\ensuremath{\mathrm{i}}\xspace}

 \def\Pp      {\ensuremath{\mathrm{p}}\xspace}

}
{

 \def\Pmu         {\ensuremath{\mu}\xspace}

 \def\Ptau        {\ensuremath{\tau}\xspace}

 \mathchardef\PDelta="7101
 \mathchardef\PXi="7104
 \mathchardef\PLambda="7103
 \mathchardef\PSigma="7106
 \mathchardef\POmega="710A
 \mathchardef\PUpsilon="7107
                  
 \def\PB      {\ensuremath{B}\xspace}                 
                  
 \def\PD      {\ensuremath{D}\xspace}

 \def\PK      {\ensuremath{K}\xspace}

 \def\PW      {\ensuremath{W}\xspace}

 \def\PZ      {\ensuremath{Z}\xspace}                 
                  
 \def\Pb      {\ensuremath{b}\xspace}                 
 \def\Pc      {\ensuremath{c}\xspace}

 \def\Pi      {\ensuremath{i}\xspace}

 \def\Pp      {\ensuremath{p}\xspace}

}

\def\mup        {\ensuremath{\Pmu^+}\xspace}
\def\mun        {\ensuremath{\Pmu^-}\xspace} 

\def\taup       {\ensuremath{\Ptau^+}\xspace}
\def\taum       {\ensuremath{\Ptau^-}\xspace}

\def\cquark    {\ensuremath{\Pc}\xspace}

\def\bquark    {\ensuremath{\Pb}\xspace}

  \def\Kbar  {\kern 0.2em\overline{\kern -0.2em \PK}{}\xspace}

  \def\Dbar    {\kern 0.2em\overline{\kern -0.2em \PD}{}\xspace}

\def\Bbar    {\ensuremath{\kern 0.18em\overline{\kern -0.18em \PB}{}}\xspace}

  \def\Y#1S{\ensuremath{\PUpsilon{(#1S)}}\xspace}

\def\Lbar {\ensuremath{\kern 0.1em\overline{\kern -0.1em\PLambda}}\xspace}

\def\AT#1     {\ensuremath{A_{\mathrm{T}}^{#1}}\xspace}           

\def\C#1      {\ensuremath{\mathcal{C}_{#1}}\xspace}                       
\def\Cp#1     {\ensuremath{\mathcal{C}_{#1}^{'}}\xspace}                    
\def\Ceff#1   {\ensuremath{\mathcal{C}_{#1}^{\mathrm{(eff)}}}\xspace}        
\def\Cpeff#1  {\ensuremath{\mathcal{C}_{#1}^{'\mathrm{(eff)}}}\xspace}       
\def\Ope#1    {\ensuremath{\mathcal{O}_{#1}}\xspace}                       
\def\Opep#1   {\ensuremath{\mathcal{O}_{#1}^{'}}\xspace}                    

\newcommand{\tev}{\ifthenelse{\boolean{inbibliography}}{\ensuremath{~T\kern -0.05em eV}\xspace}{\ensuremath{\mathrm{\,Te\kern -0.1em V}}\xspace}}
\newcommand{\gev}{\ensuremath{\mathrm{\,Ge\kern -0.1em V}}\xspace}
\newcommand{\mev}{\ensuremath{\mathrm{\,Me\kern -0.1em V}}\xspace}
\newcommand{\kev}{\ensuremath{\mathrm{\,ke\kern -0.1em V}}\xspace}
\newcommand{\ev}{\ensuremath{\mathrm{\,e\kern -0.1em V}}\xspace}
\newcommand{\gevc}{\ensuremath{{\mathrm{\,Ge\kern -0.1em V\!/}c}}\xspace}
\newcommand{\mevc}{\ensuremath{{\mathrm{\,Me\kern -0.1em V\!/}c}}\xspace}
\newcommand{\gevcc}{\ensuremath{{\mathrm{\,Ge\kern -0.1em V\!/}c^2}}\xspace}
\newcommand{\gevgevcccc}{\ensuremath{{\mathrm{\,Ge\kern -0.1em V^2\!/}c^4}}\xspace}
\newcommand{\mevcc}{\ensuremath{{\mathrm{\,Me\kern -0.1em V\!/}c^2}}\xspace}

\def\mum  {\ensuremath{\,\upmu\rm m}\xspace}

\def\gsim{{~\raise.15em\hbox{$>$}\kern-.85em
          \lower.35em\hbox{$\sim$}~}\xspace}
\def\lsim{{~\raise.15em\hbox{$<$}\kern-.85em
          \lower.35em\hbox{$\sim$}~}\xspace}

\def\evtgen     {\mbox{\textsc{EvtGen}}\xspace}
\def\fewz       {\mbox{\textsc{Fewz}}\xspace}

\def\geant      {\mbox{\textsc{Geant4}}\xspace}

\def\herwigpp     {\mbox{\textsc{Herwig++}}\xspace}

\def\photos     {\mbox{\textsc{Photos}}\xspace}

\def\pythia     {\mbox{\textsc{Pythia}}\xspace}

\def\tell1  {TELL1\xspace}
\def\ukl1   {UKL1\xspace}

\usepackage{cite} 
\usepackage{mciteplus}

\usepackage{longtable} 

\begin{document}

\renewcommand{\thefootnote}{\fnsymbol{footnote}}
\setcounter{footnote}{1}


\begin{titlepage}
\pagenumbering{roman}

\vspace*{-1.5cm}
\centerline{\large EUROPEAN ORGANIZATION FOR NUCLEAR RESEARCH (CERN)}
\vspace*{1.5cm}
\hspace*{-0.5cm}
\begin{tabular*}{\linewidth}{lc@{\extracolsep{\fill}}r}
\vspace*{-2.9cm}\mbox{\!\!\!\includegraphics[width=.14\textwidth]{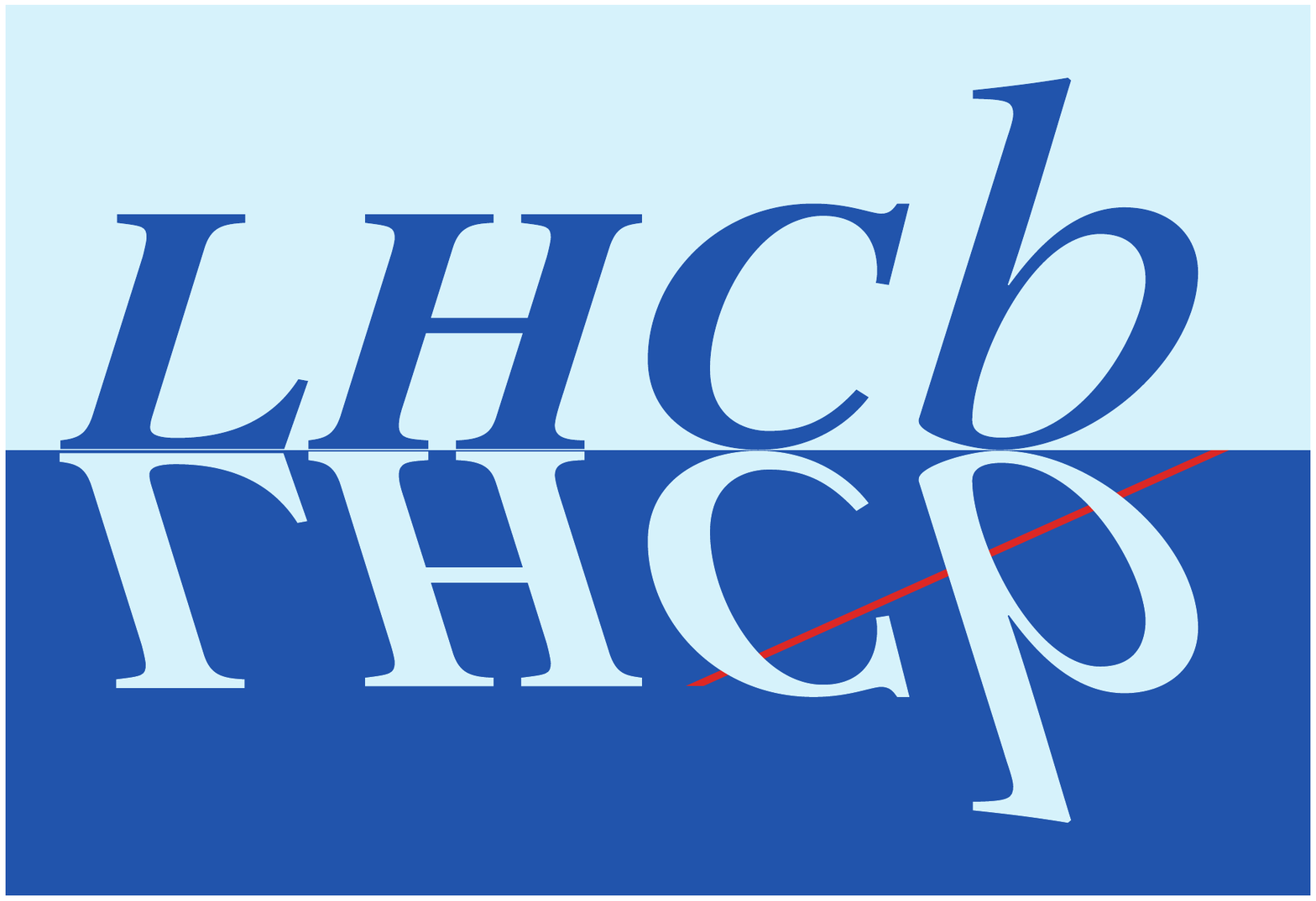}} & &%

\\
 & & CERN-PH-EP-2013-198 \\  
 & & LHCb-PAPER-2013-058 \\  
 & & December 4, 2013 \\ 
\end{tabular*}

\vspace*{2.5cm}

{\bf\boldmath\huge
\begin{center}
  Study of forward Z+jet production in pp collisions at $\sqrt{s} = 7\,\text{TeV}$
\end{center}
}

\vspace*{1.5cm}

\begin{center}
The LHCb collaboration\footnote{Authors are listed on the following pages.}
\end{center}

\vspace{\fill}

\begin{abstract}
  \noindent A measurement of the $\PZ(\rightarrow\mup\mun)+\text{jet}$ production cross-section in $\Pp\Pp$ collisions at a centre-of-mass energy $\sqrt{s} = 7\text{ TeV}$ is presented. The analysis is based on an integrated luminosity of 1.0$\text{ fb}^{-1}$ recorded by the LHCb experiment. Results are shown with two jet transverse momentum thresholds, 10 and 20~GeV, for both the overall cross-section within the fiducial volume, and for six differential cross-section measurements. The fiducial volume requires that both the jet and the muons from the $\PZ$ boson decay are produced in the forward direction ($2.0<\eta<4.5$). The results show good agreement with theoretical predictions at the second-order expansion in the coupling of the strong interaction.
\end{abstract}

\vspace*{1.5cm}

\begin{center}
  Submitted to JHEP
\end{center}

\vspace{\fill}

{\footnotesize 
\centerline{\copyright~CERN on behalf of the \lhcb collaboration, license \href{http://creativecommons.org/licenses/by/3.0/}{CC-BY-3.0}.}}
\vspace*{2mm}

\end{titlepage}


\newpage
\setcounter{page}{2}
\mbox{~}
\newpage

\centerline{\large\bf LHCb collaboration}
\begin{flushleft}
\small
R.~Aaij$^{40}$, 
B.~Adeva$^{36}$, 
M.~Adinolfi$^{45}$, 
C.~Adrover$^{6}$, 
A.~Affolder$^{51}$, 
Z.~Ajaltouni$^{5}$, 
J.~Albrecht$^{9}$, 
F.~Alessio$^{37}$, 
M.~Alexander$^{50}$, 
S.~Ali$^{40}$
% ...

{\footnotesize \it
% ...
$ ^{5}$Clermont Universit\'{e}, Universit\'{e} Blaise Pascal, CNRS/IN2P3, LPC, Clermont-Ferrand, France\\
$ ^{6}$CPPM, Aix-Marseille Universit\'{e}, CNRS/IN2P3, Marseille, France\\
% ...
$ ^{9}$Fakult\"{a}t Physik, Technische Universit\"{a}t Dortmund, Dortmund, Germany\\
% ...
$ ^{36}$Universidad de Santiago de Compostela, Santiago de Compostela, Spain\\
$ ^{37}$European Organization for Nuclear Research (CERN), Geneva, Switzerland\\
% ...
$ ^{40}$Nikhef National Institute for Subatomic Physics, Amsterdam, The Netherlands\\
% ...
$ ^{45}$H.H. Wills Physics Laboratory, University of Bristol, Bristol, United Kingdom\\
% ...
$ ^{50}$School of Physics and Astronomy, University of Glasgow, Glasgow, United Kingdom\\
$ ^{51}$Oliver Lodge Laboratory, University of Liverpool, Liverpool, United Kingdom\\
% ...
$ ^{s}$Scuola Normale Superiore, Pisa, Italy\\
}
\end{flushleft}

\cleardoublepage


\renewcommand{\thefootnote}{\arabic{footnote}}
\setcounter{footnote}{0}


\pagestyle{plain} 
\setcounter{page}{1}
\pagenumbering{arabic}


\section{Introduction}
\label{sec:Introduction}
 
Measurements of electroweak boson production in the forward region are sensitive to parton distribution functions (PDFs) at low Bjorken-$x$ which are not particularly well constrained by previous results~\cite{Thorne:2008am}. The LHCb experiment has recently presented measurements of inclusive $\PW$ and $\PZ$ boson\footnote{Throughout this article $\PZ$ includes both the $\PZ$ and the virtual photon ($\gamma^*$) contribution.} production in the muon decay channels~\cite{LHCb-PAPER-2012-008} and inclusive $\PZ$ boson production in the electron~\cite{LHCb-PAPER-2012-036} and the tau lepton~\cite{LHCb-PAPER-2012-029} decay channels. This article presents a measurement of the inclusive Z+jet production cross-section in proton-proton collisions at LHCb. These interactions typically involve the collision of a sea quark or gluon with a valence quark, and measurements of $\PZ$ boson production in association with jets are sensitive to the gluon content of the proton~\cite{JuanTalk}. LHCb is sensitive to a region of phase space in which both the $\PZ$ boson and the jet are produced in the forward region. Measurements at LHCb are therefore complementary to those at ATLAS~\cite{Aad:2013ysa} and CMS~\cite{Chatrchyan:2013tna,Chatrchyan:2013oda}. Hence, measurements of the $\PZ$+jet production cross-section at LHCb enable comparisons of different PDF predictions and their relative performances in this previously unprobed region of phase space.
 
The $\PZ$+jet production cross-section, in addition to being sensitive to the PDFs at low Bjorken-$x$, is influenced by higher order contributions in perturbative quantum chromodynamics (pQCD). Studies of the Drell-Yan process in the forward region are sensitive to multiple radiation of partons~\cite{Hautmann:2012sh}. Measurements in the forward region have not been used to tune generators and, consequently, studies of Z+jet production in the forward region can be used to test the accuracy of different models. Theoretical predictions for the Z+jet process are available at $\mathcal{O}(\alpha_s^2)$~\cite{Arnold:1988dp,Giele:1993dj,Anastasiou:2003ds,Gavin:2010az,Campbell:2010ff,PowhegZj,Mangano:2002ea,Gleisberg:2008ta}, where $\alpha_s$ is the strong-interaction coupling strength. Similar analyses at ATLAS~\cite{Aad:2013ysa} and CMS~\cite{Chatrchyan:2013tna,Chatrchyan:2013oda} have shown reasonable agreement between data and such predictions.

This measurement of the cross-section of $\PZ\rightarrow\mup\mun$ events with jets in the final state uses data corresponding to an integrated luminosity of $1.0\,\rm{fb}^{-1}$ taken by the LHCb experiment in $\Pp\Pp$ collisions at a centre-of-mass energy of 7~TeV. The analysis is performed in a fiducial region that closely corresponds to the kinematic coverage of the LHCb detector. For the dimuon decay of the $\PZ$ boson, this requirement is the same as that in Ref.~\cite{LHCb-PAPER-2012-008}. Both final state muons are required to have a transverse momentum\footnote{Throughout this article natural units, where $c=1$, are used.}, $p_\text{T}^\mu$, greater than 20~GeV, and to have pseudorapidity\footnote{The pseudorapidity is defined to be $\eta \equiv -\ln (\tan (\theta/2))$, where the polar angle $\theta$ is measured with respect to the beam axis. The rapidity of a particle is defined to be $y \equiv 0.5 \ln[(E+p_z)/(E-p_z)]$, where the particle has energy $E$ and momentum $p_z$ in the direction of the beam axis.} in the range $2.0<\eta^\mu <4.5$. The invariant mass of the dimuon system is required to be in the range $60<M_{\mu\mu}<120$~GeV. Jets are reconstructed using the anti-$k_\text{T}$ algorithm~\cite{Cacciari:2008gp} with distance parameter $R = 0.5$, and are required to be in the fiducial region $2.0<\eta^\text{jet}<4.5$, and to be separated from decay muons of the $\PZ$ boson by $\Delta r(\mu, \text{jet})>0.4$. This separation is defined such that $\Delta r^2 \equiv \Delta\phi^2 + \Delta\eta^2$, where $\Delta\phi$ is the difference in azimuthal angle and $\Delta\eta$ the difference in pseudorapidity between the muon and the jet directions. Results are presented for two thresholds of the jet transverse momentum: $p_\text{T}^\text{jet}>20$~GeV and $p_\text{T}^\text{jet}>10$~GeV. Both the total $\PZ$+jet cross-section and the cross-section ratio of \PZ+jet production to inclusive \PZ production are reported. In addition, six differential cross-sections for $\PZ$+jet production are presented as a function of the $\PZ$ boson rapidity and transverse momentum, the pseudorapidity and transverse momentum of the leading\footnote{The leading jet is defined to be the highest transverse momentum jet in the fiducial region.} jet, and the difference in azimuthal angle and in rapidity between the $\PZ$ boson and this jet. These differential measurements are presented normalised to the total $\PZ$+jet cross-section. The data are compared to predictions at $\mathcal{O}(\alpha_s)$ and $\mathcal{O}(\alpha_s^2)$ using different PDF parametrisations.

The remainder of this article is organised as follows: Sect.~\ref{sec:Detector} describes the LHCb detector and the simulation samples used; Sect.~\ref{sec:Jetreco} provides an overview of jet reconstruction at LHCb; Sect.~\ref{sec:Selection} describes the selection and reconstruction of candidates and the determination of the background level; Sect.~\ref{sec:Method} describes the cross-section measurement; the associated systematic uncertainties are discussed in Sect.~\ref{sec:Systematics}; the results are presented in Sect.~\ref{sec:Results}; Sect.~\ref{sec:Conclusion} concludes the article.

\section{LHCb detector and simulation}
\label{sec:Detector}

The \lhcb detector~\cite{Alves:2008zz} is a single-arm forward
spectrometer covering the \mbox{pseudorapidity} range $2<\eta <5$,
designed for the study of particles containing \bquark or \cquark
quarks. The detector includes a high-precision tracking system
consisting of a silicon-strip vertex detector (the VELO) surrounding the $\Pp\Pp$
interaction region, a large-area silicon-strip detector (the TT) located
upstream of a dipole magnet with a bending power of about
$4{\rm\,Tm}$, and three stations of silicon-strip detectors and straw
drift tubes placed downstream.
The combined tracking system provides a momentum measurement with
relative uncertainty that varies from 0.4\,\% at 5\gev to 0.6\,\% at 100\gev,
and impact parameter resolution of 20\mum for
tracks with large transverse momentum. Charged hadrons are identified
using two ring-imaging Cherenkov detectors~\cite{LHCb-DP-2012-003}. Photon, electron and
hadron candidates are identified by a calorimeter system consisting of
scintillating-pad (SPD) and preshower detectors, an electromagnetic
calorimeter and a hadronic calorimeter. Muons are identified by a
system composed of alternating layers of iron and multiwire
proportional chambers~\cite{LHCb-DP-2012-002}.
The trigger~\cite{LHCb-DP-2012-004} consists of a
hardware stage, based on information from the calorimeter and muon
systems, followed by a software stage, which applies a full event
reconstruction.

To avoid the
possibility that a few events with high occupancy dominate the CPU time of the software
trigger, a set of global event cuts (GEC) is applied on the hit multiplicities of most subdetectors
used in the pattern recognition algorithms. The dominant GEC in the trigger selection used in this analysis is the requirement that the hit
multiplicity in the SPD, $n_\text{SPD}$, is less than 600.

In the simulation, $\Pp\Pp$ collisions are generated using
\pythia~6.4~\cite{Sjostrand:2006za} with a specific \lhcb
configuration~\cite{LHCb-PROC-2010-056}, with the CTEQ6ll~\cite{Nadolsky:2008zw} parametrisation for the PDFs. 
Decays of hadronic particles
are described by \evtgen~\cite{Lange:2001uf}, in which final state
radiation is generated using \photos~\cite{Golonka:2005pn}. The
interaction of the generated particles with the detector and its
response are implemented using the \geant
toolkit~\cite{Allison:2006ve, *Agostinelli:2002hh} as described in
Ref.~\cite{LHCb-PROC-2011-006}. The main simulation sample used in this analysis is an $\mathcal{O}(\alpha_s)$ prediction of the $\PZ$+jet process, with the $\PZ$ boson decaying to two muons. In addition, inclusive $\PZ\rightarrow\mup\mun$ events are generated at leading order in pQCD, where all jets are produced by the parton shower, in order to study various stages of the analysis with an independent simulation sample. This simulation sample is hereafter referred to as the inclusive $\PZ$ sample.

\section{Jet reconstruction}
\label{sec:Jetreco}
Inputs for jet reconstruction are selected using a particle flow algorithm. In order to benefit from the good momentum resolution of the LHCb tracking system, reconstructed tracks serve as charged particle inputs to the jet reconstruction. Tracks corresponding to the decay muons of the Z boson are excluded. The neutral particle inputs are derived from the energy deposits in the electromagnetic and hadronic calorimeters. If the deposits are matched to tracks, the expected calorimeter energies associated with the tracks are subtracted. The expected calorimeter energy is determined based on the likelihood that the track is associated with a charged hadron, a muon, or an electron, using information from the particle identification systems. If a significant energy deposit remains after the subtraction, the energy is associated with a neutral particle detected in the calorimeter. The use of the different particle identification hypotheses has negligible impact on the results presented in this article, since the jets studied here are mostly inititated by light quarks and gluons.
Finally, in order to reduce the contribution from multiple proton-proton interactions, charged particles from tracks reconstructed within the VELO are not considered if they are associated to a different primary vertex to that of the $\PZ$ boson. The charged particles and energy clusters are reconstructed into jets using the anti-$k_\text{T}$ algorithm~\cite{Cacciari:2008gp}, with distance parameter $R = 0.5$, as implemented in {\sc Fastjet}~\cite{Cacciari:2005hq}.

The same jet reconstruction algorithm is run on simulated Z+jet events. The anti-$k_\text{T}$ algorithm is also applied to these simulated events at the hadron-level using information that is available before the detector simulation is performed. The inputs for these `true' jets are all stable final state particles, including neutrinos, from the same proton-proton interaction that produced the $\PZ$ boson, that are not products of the $\PZ$ boson decay.

The transverse momentum of a reconstructed jet is scaled so that it gives an unbiased estimate of the true jet transverse momentum. The scaling factor, typically between $0.9$ and $1.1$, is determined from simulation and depends on the jet pseudorapidity and transverse momentum, the fraction of the jet transverse momentum measured with the tracking systems, and the number of proton-proton interactions in the event. The energy resolution of reconstructed jets varies with the jet energy. The half width at half maximum for the distribution of $p_\text{T}^\text{reco} / p_\text{T}^\text{true}$ is typically 10-15\,\% for jets with transverse momenta between 10 and 100~GeV. In simulation, 90\,\% of jets with at least 10~GeV transverse momentum are reconstructed with $\Delta r < 0.13$ in $\eta-\phi$ space with respect to the true jet. At the $p_\text{T}$ threshold of 20~GeV the corresponding radius is 0.08.

In order to reduce the number of spurious fake jets, and to select jets from the same interaction as that of the $\PZ$ boson with a good estimate of the jet energy, additional jet identification requirements are imposed. Jets are required to contain at least two particles matched to the same primary vertex, to contain at least one track with $p_\text{T}>1.8$~GeV, and to contain no single particle with more than $75\,\%$ of the jet's transverse momentum.

\section{Selection and event reconstruction}
\label{sec:Selection}
The $\PZ\rightarrow\mup\mun$ selection follows that described in Ref.~\cite{LHCb-PAPER-2012-008}. The events are initially selected by a trigger that requires the presence of at least one muon candidate with $p_\text{T}^\Pmu>10~\text{GeV}$. Selected events are required to contain two reconstructed muons with $p_\text{T}^\mu>20~\text{GeV}$ and $2.0<\eta^\mu<4.5$, and one of these muons is required to have passed the trigger. The invariant mass of the dimuon pair must be in the range $60<M_{\mu\mu}<120\text{ GeV}$. The relative uncertainty on the measured momentum of each muon is required to be less than $10\,\%$ and the $\chi^2$ probability for the associated track larger than $0.1\,\%$. In total, $53\,182$ $\PZ\rightarrow\mup\mun$ candidates are selected.

A reconstructed jet with pseudorapidity in the range $2.0<\eta^\text{jet}<4.5$ is also required in the selection. The separation between each of the decay muons of the reconstructed $\PZ$ boson and the jet is required to be $\Delta r>0.4$. Jets are reconstructed with transverse momentum above 7.5~GeV. Of the selected $\PZ\rightarrow\mup\mun$ candidates, $4\,118$ contain a reconstructed jet with transverse momentum above 20~GeV, and $10\,576$ contain a jet with transverse momentum above 10~GeV.

\subsection{Background}
\label{subs:BKG}
The background contribution from random combinations of muons can come from semileptonic heavy flavour decays, $\PW$ boson decays, or mesons that have decayed whilst passing through the detector and have been reconstructed as muons, or hadrons that have passed through the calorimeters without interacting. This background is determined from the number of events containing two muons of the same charge that would otherwise pass the selection requirements. No significant difference is found using events where both muons have a positive charge or events where both muons have a negative charge. This background source contributes $5\pm2$ events for the 20~GeV jet transverse momentum threshold and $16\pm4$ for the 10~GeV threshold, where the uncertainties are statistical. The production of diboson pairs and heavy flavour decays of $\PZ$ bosons, where the heavy flavour decay products decay to muons, are found to contribute negligible background levels to this analysis.

Decays from the $\PZ\rightarrow\taup\taum$ process where both tau leptons decay to muons and neutrinos are another potential background source. This background is determined from simulation, and contributes $7\pm3$ events for the 20~GeV transverse momentum threshold, and $12\pm3$ events for the 10~GeV threshold, where the uncertainties are statistical.

The background contribution from top quark pair production is also considered, where the top quark decay products include high transverse momentum muons. This background is determined from next-to-leading order (NLO) simulation to be $5\pm2$ events, where the uncertainties are statistical. This background is largely independent of the 10 and 20~GeV jet transverse momentum thresholds as the top quark decays are associated with very high transverse momentum jets.

The background associated with events where a jet above a threshold is reconstructed, despite there being no true jet above that threshold, is treated as a migration. This background is corrected for by unfolding the transverse momentum distribution (see Sect.~\ref{sec:Method}).

The total background contribution for the 20~GeV jet transverse momentum threshold is $17\pm4$ events, and the contribution for the 10~GeV threshold is $33\pm6$ events. This corresponds to a sample purity $\rho \equiv S/(S+B)$, where $S$ is the number of signal events and $B$ is the number of background events, of $(99.6\pm0.1)\,\%$ for the 20~GeV threshold and $(99.7\pm0.1)\,\%$ for the 10~GeV threshold. These purities are consistent with that found in the inclusive $\PZ$ boson analysis~\cite{LHCb-PAPER-2012-008}. The purity shows no significant dependence on other kinematic variables of interest. Since the purity is high and has little variation with the transverse momentum threshold it is treated as constant for this analysis.
\subsection{Z detection and reconstruction efficiencies}
\label{subs:ZDetEps}
Following Ref.~\cite{LHCb-PAPER-2012-036}, the total Z boson detection efficiency is factorised into four separate components as $\varepsilon_{\PZ} = \varepsilon_\text{GEC}\, \varepsilon_\text{trigger}\,  \varepsilon_\text{track}\,  \varepsilon_\text{ID}$, where the $\varepsilon_\text{X}$ factors correspond to the efficiency associated with the GEC, the trigger requirements, the muon track reconstruction and the muon identification, respectively.                                

The GEC, applied in the trigger to stop very large events dominating processing time, cause signal events to be rejected. The associated inefficiency is obtained using the same method described in Ref.~\cite{LHCb-PAPER-2012-036}, where an alternative dimuon trigger requirement\footnote{This trigger route is not used elsewhere in this analysis as it has a lower efficiency than the single muon trigger.} is used to determine the number of events that are rejected with $600<n_\text{SPD}<900$. The small number of events with $n_\text{SPD}>900$ is found by extrapolation using a fit with a gamma function. This approach is applied to determine the efficiency as a function of the number of reconstructed primary vertices and the number of jets reconstructed in the event. The average efficiency is $91\,\%$.

The trigger efficiency for the single muon trigger is found using the same tag-and-probe method used in Ref.~\cite{LHCb-PAPER-2012-008}. Events in which at least one muon from the $\PZ$ boson decay passed the trigger are selected. The fraction of events where the other muon from the $\PZ$ boson decay fired the trigger determines the muon trigger efficiency. This efficiency is found to be independent of the number of jets reconstructed in the event and is determined as a function of the muon pseudorapidity. The efficiencies for the two muons are then combined to determine the efficiency with which at least one of the two muons in the decay passes the trigger, $\varepsilon_\text{trigger}(\eta_1, \eta_2) = \varepsilon(\eta_1)+\varepsilon(\eta_2) - \varepsilon(\eta_1)\varepsilon(\eta_2)$. This combination assumes that the probability that one muon fires the trigger is independent of whether the other muon fired the trigger. This is confirmed with simulated data. The average of this combined efficiency is approximately $96\,\%$.

The muon track reconstruction efficiency is determined using the tag-and-probe method, described in Refs.~\cite{LHCb-PAPER-2012-008} and~\cite{Jaeger:1402577}. Well reconstructed tracks in the muon stations are linked to hits in the TT detector in events containing one other high-purity muon candidate. The invariant mass of this dimuon pair is required to lie within 10~GeV of the $\PZ$ boson mass. The efficiency is determined as the fraction of events where the muon-station track is geometrically matched to a track in the tracking system that passes the track quality requirements. This efficiency depends on the muon pseudorapidity and the number of jets measured in the event, with an average efficiency of approximately $90\,\%$ for each muon.

The muon identification efficiency is determined using the method described in  Ref.~\cite{LHCb-PAPER-2012-008}. Events containing two tracks with an invariant mass within 5~GeV of the $\PZ$ boson mass are selected. One of the tracks is required to be identified as a muon. The fraction of events in which the other track is also identified as a muon defines the muon identification efficiency. This efficiency shows no dependence on the number of jets in the event and is found as a function of the muon pseudorapidity. The average muon identification efficiency is $99\,\%$.

\subsection{Jet detection and reconstruction efficiencies}
\label{subs:JetDetEps}
\begin{figure}[h]
  \centering  
  \includegraphics[width=0.75\textwidth]{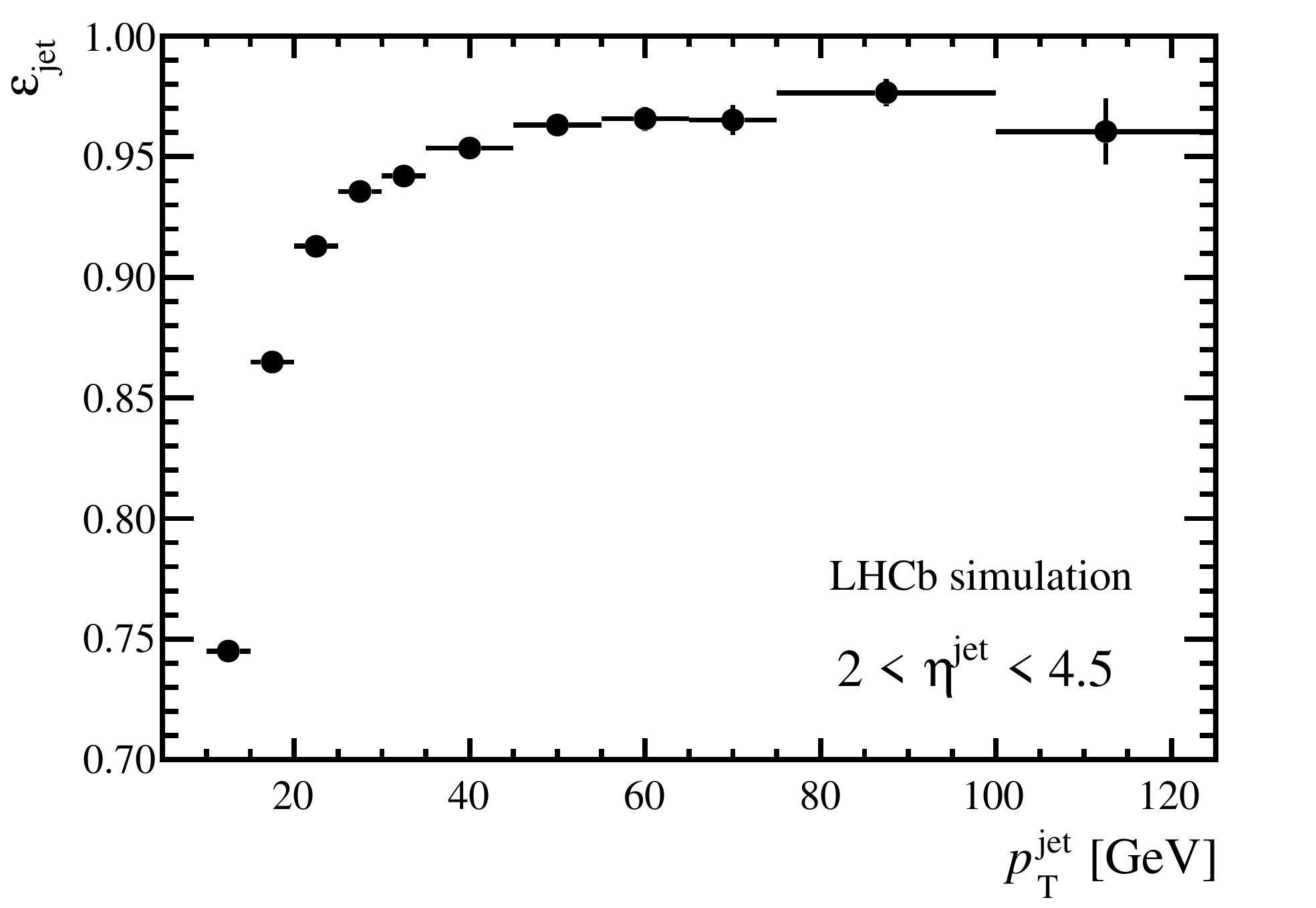}
  \caption{\small Jet identification efficiency as a function of the true jet $p_\text{T}$. The uncertainties shown are statistical. The zero on the vertical axis is suppressed.\normalsize}
  \label{IDeff}
\end{figure}

The jet detection efficiency is determined from simulation and is defined as the efficiency for a jet to be reconstructed with transverse momentum greater than 7.5~GeV, satisfying the jet selection criteria, given that a true jet is reconstructed in the same event. This efficiency is determined as a function of the true jet transverse momentum and shows little variation in the central region of the LHCb detector. Reweighting the simulation to have the same jet pseudorapidity distribution as data has a negligible effect on the efficiency. This efficiency is about $75\,\%$ for jets with transverse momentum of about 10~GeV, but rises to about $96\,\%$ for high transverse momentum jets, as shown in Fig.~\ref{IDeff}. The drop in efficiency at low transverse momentum is mainly due to the jet identification requirements having a larger effect in this region. 

\section{Cross-section measurement}
\label{sec:Method} 
Events are selected with reconstructed jet transverse momentum above 7.5 GeV. Migrations in the jet transverse momentum distribution are corrected for by unfolding the distribution using the method of D'Agostini~\cite{D'Agostini}, as implemented in {\sc RooUnfold}~\cite{RooUnfold}. Two iterations are chosen as this gives the best agreement between the unfolded distribution and the true distribution when the inclusive $\PZ$ simulation sample is unfolded, using the same number of events in the inclusive $\PZ$ simulation sample as are present in data. As a cross-check, the result is compared with the SVD unfolding method~\cite{SVD}. In these studies underflow bins are included in the unfolded distributions to account for the small number of events that lie below threshold after the unfolding procedure.

Each event is assigned a weight for the Z boson reconstruction, detection and selection efficiency, $\varepsilon_{\PZ}$. This enables the determination of the fraction of events within each bin of the unfolded jet transverse momentum, $N(p_\text{T}^\text{unf})$, corrected for the Z detection efficiency
\begin{equation} 
N(p_\text{T}^\text{unf}) = \sum_{\text{events}} \frac{M(p_\text{T}^\text{unf},p_\text{T}^\text{reco})}   {\varepsilon_\PZ},
\end{equation}
where $M(p_\text{T}^\text{unf},p_\text{T}^\text{reco})$ is the element of the matrix, obtained from the unfolding, that gives the probability that an event containing a jet with reconstructed transverse momentum in the bin $p_\text{T}^\text{reco}$ contains a true jet with transverse momentum in the bin $p_\text{T}^\text{unf}$. For the differential distributions the matrix is determined for events restricted to the relevant bin in that differential distribution. This unfolding includes the correction for the background where a jet is reconstructed with $p_\text{T}$ above the threshold despite there being no true jet above that threshold in the event.
  
In order to measure the cross-section, a correction is applied to account for the jet reconstruction efficiency, $\varepsilon_\text{jet}$. The correction is performed for each bin in each differential distribution separately. In differential measurements an additional factor $A_\text{mig}$ is applied to account for migration between different bins (for example, in the jet pseudorapidity distribution). These corrections are typically small ($2-3\,\%$) and are taken from simulation. The cross-section is determined by dividing the resulting event yield, corrected for migrations and the reconstruction acceptance, by the integrated luminosity, $\int\mathcal{L}\;\text{d}t$, as follows
\begin{equation}
\sigma = \frac{\rho}{\int\mathcal{L}\;\text{d}t}\sum_{p_\text{T}^\text{unf}>p_\text{T}^\text{thr}}\frac{A_\text{mig}}{\varepsilon_\text{jet}}N(p_\text{T}^\text{unf}),
\end{equation}
where $p_\text{T}^\text{thr}$ is the relevant threshold, 20 or 10~GeV, and the sum is over the bins of the unfolded transverse momentum above this threshold. The purity of the sample, $\rho$, accounts for the presence of background as discussed in Sect.~\ref{subs:BKG}. The luminosity is determined as described in Ref.~\cite{LHCB-PAPER-2011-015}. 

Measurements of the total $\PZ$+jet cross-section are quoted at the Born level in QED; the correction factors for final state radiation (FSR) of the muons are calculated with \herwigpp~\cite{HERWIG}. Differential distributions are compared to theoretical predictions that include the effects of FSR, so they are not corrected for FSR from the muons. The differential distributions are also normalised to the total Z+jet cross-section above the relevant transverse momentum threshold, without corrections for FSR, so that their integral is unity.

\section{Systematic uncertainties}
The different contributions to the systematic uncertainty are discussed below and are summarised in Table~\ref{tab:Syst}.
\begin{table}
\caption{\small The relative uncertainty arising from each source of possible systematic uncertainties considered for the Z+jet cross-section for $p_\text{T}^\text{jet}>20\text{ GeV}$. The relative uncertainties are similar for the 10~GeV threshold. The contributions from the different sources are combined in quadrature.\normalsize}
  \centering
  \begin{center}
    \begin{tabular}{l c } 
      {Source} & {Relative uncertainty (\%)}\\ \hline 
      Unfolding & 1.5\\
      Z detection and reconstruction   & 3.5\\ 
      Jet-energy scale, resolution and reconstruction & 7.8\\ 
      Final state radiation & 0.2\\\hline      
      Total excluding luminosity & 8.6  \\\hline 
      Luminosity & 3.5\\
    \end{tabular}
    \end{center}
    \label{SystSummary}
\label{tab:Syst}
\end{table}  
\label{sec:Systematics}

Two contributions associated with the unfolding are considered. The difference in the unfolded result between the SVD~\cite{SVD} and the D'Agostini~\cite{D'Agostini} methods is assigned as an uncertainty. In addition, the unfolding process is carried out on the inclusive $\PZ$ sample described in Sect.~\ref{sec:Detector} (which is an independent simulation sample to that used to perform the unfolding), and the difference between the unfolded distribution and the true distribution is assigned as a systematic uncertainty. The number of events considered in the independent sample is the same as the number in data. The differences between the results found using the D'Agostini method with one iteration and those found using two iterations are less than the uncertainties assigned from the unfolding method.

The systematic uncertainties for the muon identification and trigger efficiencies are obtained as in Ref.~\cite{LHCb-PAPER-2012-008}, where the statistical uncertainties on the tag-and-probe method are used as systematic uncertainties on the efficiency. The systematic uncertainty associated with the GEC efficiency is considered as in Ref.~\cite{LHCb-PAPER-2012-036}. A variation in the fit model is applied and the change in efficiency is considered as a systematic uncertainty. In addition, the statistical uncertainty in the efficiency is assigned as a systematic uncertainty. The systematic uncertainty associated with the track reconstruction efficiency has two contributions. The uncertainty associated with the statistical precision of the efficiency determination is treated as in Ref.~\cite{LHCb-PAPER-2012-008}. By comparing the tag-and-probe method applied to simulation with the true efficiency, the method is found to be accurate to $0.3\,\%$ for each muon. This sets the systematic uncertainty associated with the tag-and-probe method used to find the muon track reconstruction efficiency.

\begin{figure}[b]
  \centering  
  \includegraphics[width=0.7\textwidth]{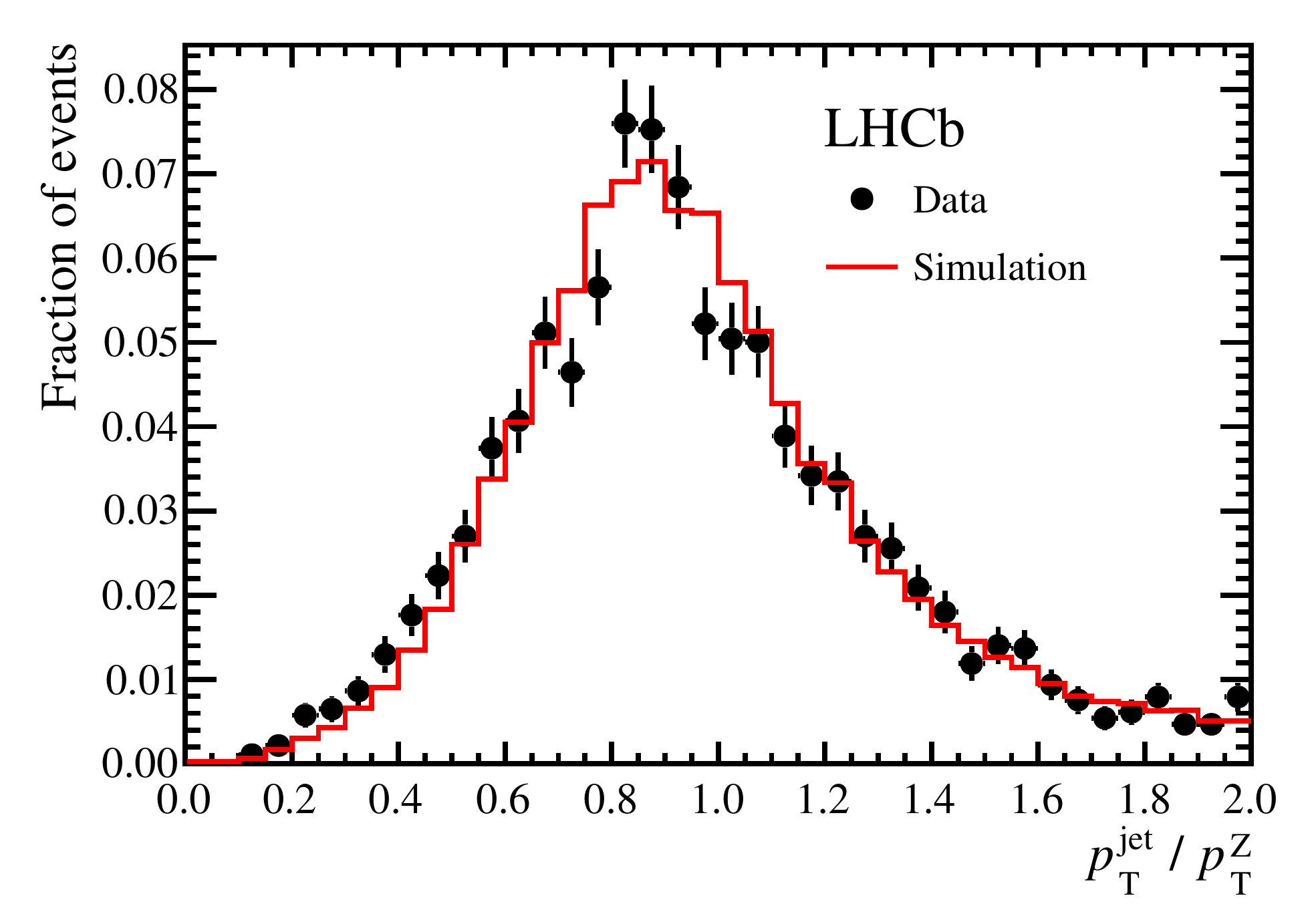}
  \caption{Comparison between data (black points) and simulation (red line) in the $p_\text{T}^\text{jet} / p_\text{T}^{\PZ}$ distribution for selected $\PZ$+1-jet events where the $\PZ$ boson and the jet are emitted azimuthally opposed. The uncertainties shown are statistical.}
  \label{Agreement}
\end{figure} 

The systematic uncertainty associated with the jet identification requirements is determined by tightening these requirements and comparing the fraction of events rejected in data and simulation. These are found to agree at the level of about $3\,\%$. This is therefore used as a systematic uncertainty. The efficiency is cross-checked on the independent inclusive $\PZ$ sample, and the difference is taken as an additional systematic uncertainty. The efficiency associated with the jet reconstruction, neglecting the jet identification requirements, is found to be about $98.5\,\%$ at low transverse momentum, so an additional $1.5\,\%$ uncertainty is assigned to this reconstruction efficiency component of $\varepsilon_\text{jet}$ at low momentum. The jet-energy scale and resolution show no dependence on the separation of the Z and the jet in the azimuthal angle. The jet-energy scale and resolution uncertainties associated with how well the detector response to jets is modelled in simulated data are therefore determined by selecting $\PZ$+1-jet events that are azimuthally opposed. In these events the $\PZ$ boson and jet transverse momenta are expected to balance. Hence, the $\PZ$ boson transverse momentum can be used as a proxy for the true jet transverse momentum. The $p_\text{T}^\text{jet}/p_\text{T}^\PZ$ distribution in the selected events is shown in Fig.~\ref{Agreement}, and is also considered as a function of the jet pseudorapidity and transverse momentum. The mean is found to agree between data and simulation at the level of about $3\,\%$, consistent within the statistical precision. The width is consistent between data and simulation, and the resolution in simulation can be smeared at the level of about $10\,\%$ whilst maintaining this agreement. Based on these comparisons, systematic uncertainties to account for the reliability of the modelling are assigned to the jet-energy scale and resolution. In addition, a systematic uncertainty is assigned based on the difference in the jet-energy scale for gluon- and quark-initiated jets, and for the method used to correct the jet-energy scale. This contributes an additional $2\,\%$ systematic uncertainty on the jet-energy scale. These uncertainties are then propagated into the cross-sections and distributions measured. The contribution from the uncertainty on the jet-energy scale is the dominant uncertainty in most bins analysed.

The systematic uncertainty on the FSR correction applied to the total cross-section is determined by comparing the correction taken from \herwigpp~\cite{HERWIG} and from {\sc Pythia}~\cite{Sjostrand:2006za} interfaced with {\sc Photos}~\cite{Golonka:2005pn}, as found in Ref.~\cite{LHCb-PAPER-2012-008}. The difference in correction is at the level of $0.2\,\%$.

The luminosity uncertainty is estimated to be $3.5\,\%$, as detailed in Ref.~\cite{LHCB-PAPER-2011-015}.


\section{Results}
\label{sec:Results}
The $\PZ$+jet cross-section and the cross-section ratio $\sigma(\text{Z+jet})/\sigma(\text{Z})$ are measured at the Born level. For the $p_\text{T}^{\text{jet}}>20\text{ GeV}$ threshold the results are
\vspace{0.2cm}

\begin{tabular}{r@{$\displaystyle\text{ }=\text{  }$ }l}
\vspace{0.2cm}
$\displaystyle\sigma(\text{Z+jet}) $ & $\displaystyle6.3\pm0.1\,(\text{{stat.}})\pm0.5\,(\text{{syst.}})\pm0.2\,(\text{{lumi.}})\text{ pb,} $ \\
\vspace{0.2cm}  
$\displaystyle\frac{\sigma(\text{Z+jet})}{\sigma(\text{Z})}$ &$ \displaystyle 0.083\pm0.001\,(\text{{stat.}})\pm0.007\,(\text{{syst.}}),  $ \\ 
\end{tabular}

\noindent and for the $\displaystyle p_\text{T}^{\text{jet}}>10\text{ GeV}$ threshold, 

\vspace{0.2cm}
\begin{tabular}{r@{$\displaystyle\text{ }=\text{  }$ }l}
\vspace{0.2cm}  
$\displaystyle\sigma(\text{Z+jet})$ & $\displaystyle16.0\pm0.2\,(\text{{stat.}})\pm1.2\,(\text{{syst.}})\pm0.6\,(\text{{lumi.}})\text{ pb,} $ \\
\vspace{0.2cm}   
$\displaystyle\frac{\sigma(\text{Z+jet})}{\sigma(\text{Z})}$ & $\displaystyle0.209\pm0.002\,(\text{{stat.}})\pm0.015\,(\text{{syst.}}),  $ 
\end{tabular}

\noindent where the first uncertainty is statistical, the second is systematic and the third is the uncertainty due to the luminosity determination.

The measured cross-sections are compared to theoretical predictions at $\mathcal{O}(\alpha_s^2)$ calculated using {\sc{Powheg}}\cite{PowhegZj,Nason:2004rx,Frixione:2007vw,Alioli:2010xd}. The parton shower development and hadronisation are simulated using {\sc{Pythia} 6.4}~\cite{Sjostrand:2006za}, with the Perugia~0 tune~\cite{Skands:2010ak}. Jets are created out of all stable particles in the final state that are not produced by the decay of the $\PZ$ boson. These predictions are computed with the renormalisation scale and factorisation scales set to the nominal value of the vector boson transverse momentum.

The theoretical predictions are computed for three different NLO PDF parametrisations: MSTW08~\cite{Martin:2009iq}, CTEQ10~\cite{Lai:2010vv} and NNPDF~2.3~\cite{Ball:2012cx}. For the differential distributions, the CTEQ10 and NNPDF~2.3 results are calculated at $\mathcal{O}(\alpha_s^2)$. Results using the MSTW08 parametrisation are calculated at $\mathcal{O}(\alpha_s)$ and $\mathcal{O}(\alpha_s^2)$. For the ratio $\sigma_{\PZ+\text{jet}} / \sigma_\PZ$, the Z+jet cross-section is computed at $\mathcal{O}(\alpha_s^2)$ and the Z cross-section at $\mathcal{O}(\alpha_s)$ for the MSTW08, CTEQ10 and NNPDF~2.3 PDF parametrisations.  To see the effect of higher orders in pQCD on the Z+jet cross-section, theoretical predictions are also computed by taking the ratio between the Z and Z+jet cross-sections at $\mathcal{O}(\alpha_s)$, with the PDFs determined from the MSTW08 NLO parametrisation.
 
In addition, the $\PZ$+jet cross-section is computed using \fewz~\cite{Gavin:2010az} at $\mathcal{O}(\alpha_s^2)$, with the MSTW08 NLO PDF parametrisation. The cross-section for inclusive $\PZ$ boson production is calculated using FEWZ at $\mathcal{O}(\alpha_s)$, with the same PDF parametrisation. This theoretical prediction neglects effects from hadronisation and the underlying event, and so comparisons with the results and the other predictions are indicative of the size of these effects.  For these calculations the renormalisation scale and factorisation scales are set to the nominal value of the vector boson mass. 
 
Uncertainties on all predictions are calculated by repeating the calculations with the renormalisation and factorisation scales simultaneously varied by a factor of two about their nominal values. The spread in predictions from the different PDF parametrisations is indicative of the PDF uncertainty.
  
\begin{figure}
  \centering   
  \includegraphics[width=0.65\textwidth]{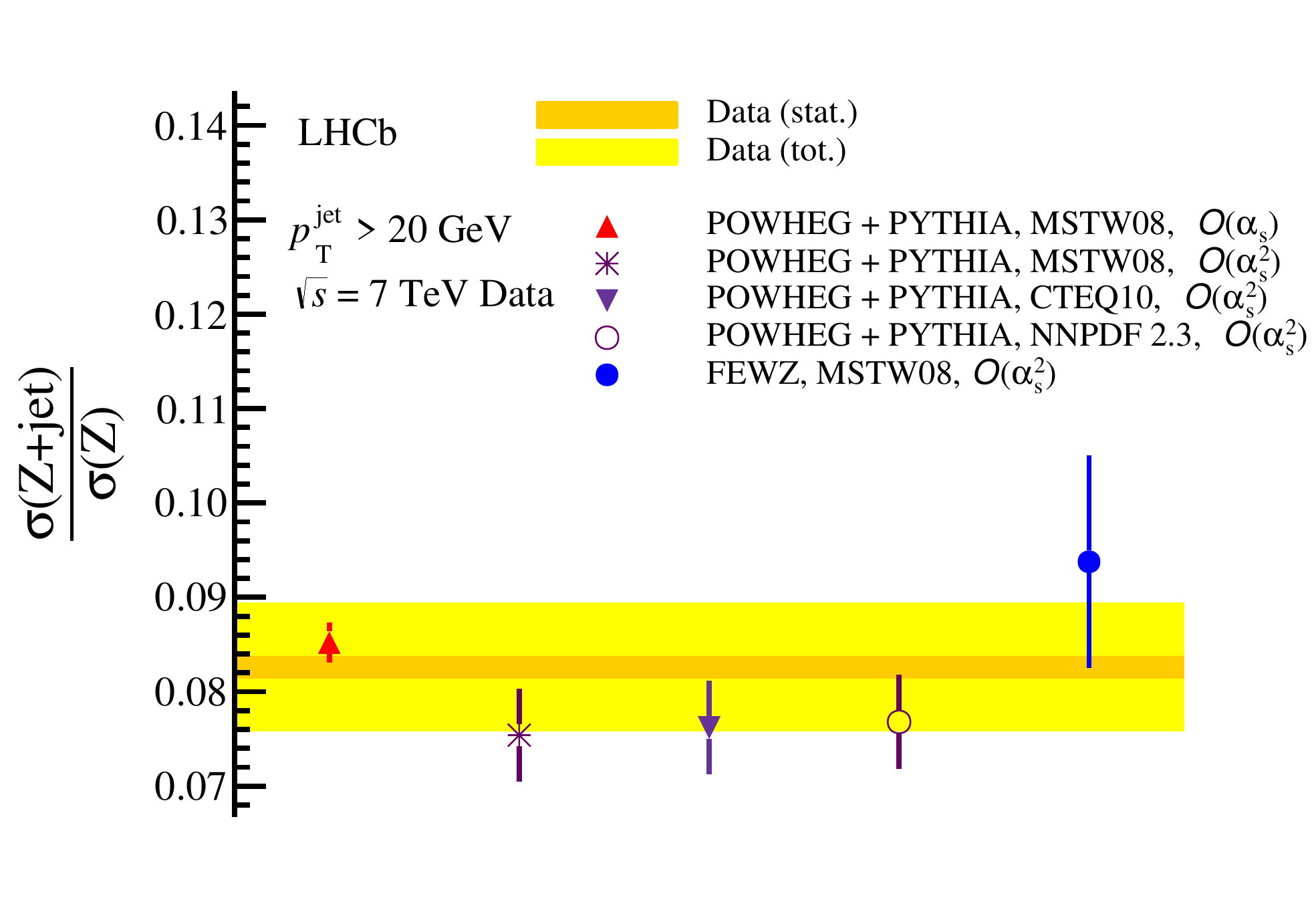}
  \includegraphics[width=0.65\textwidth]{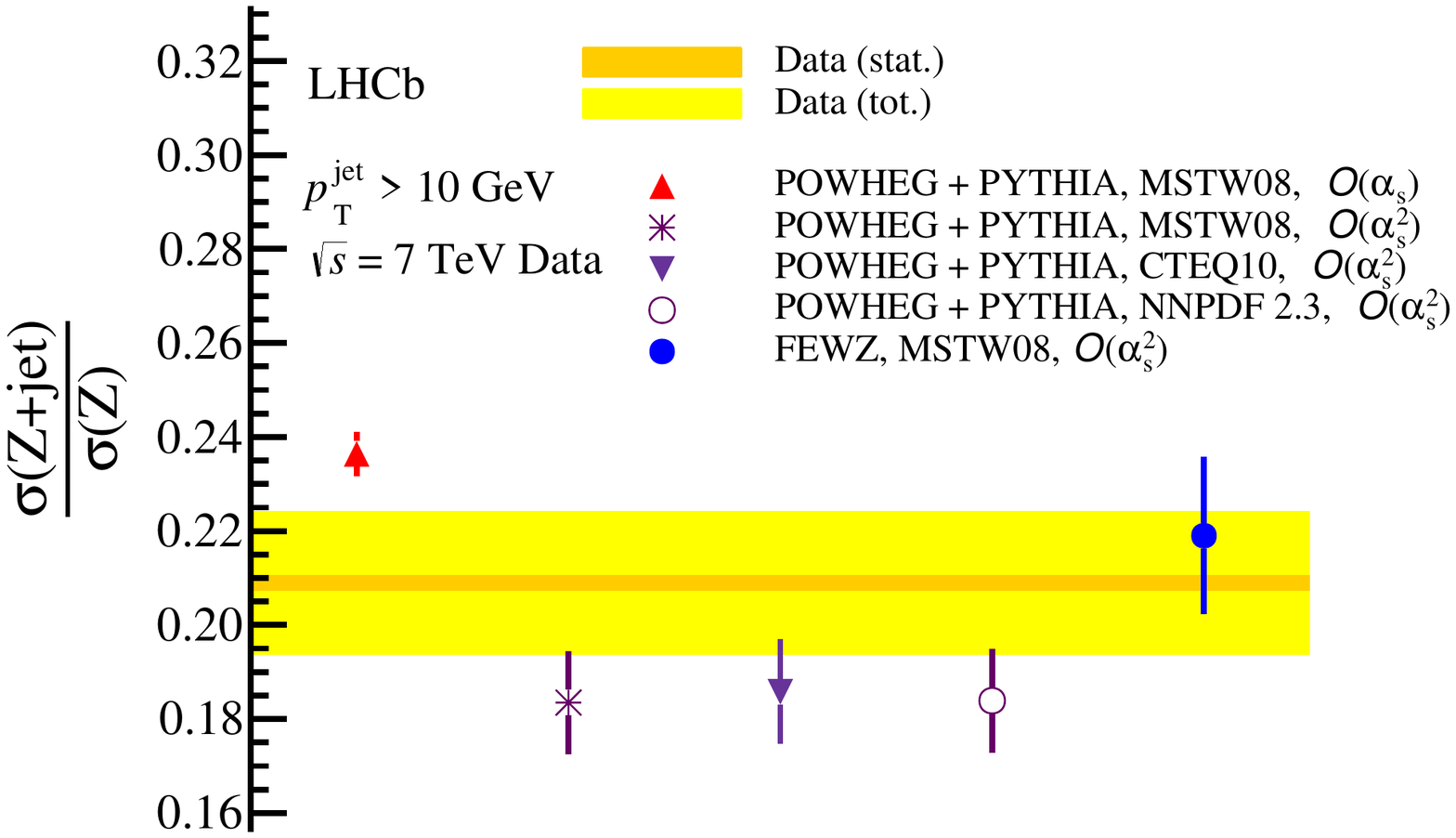}
  \caption{\small Ratio of the Z+jet cross-section to the inclusive cross-section, for (top) $p_\text{T}^\text{jet}\nobreak>\nobreak 20\text{ GeV}$ and (bottom) $p_\text{T}^\text{jet} > 10\text{ GeV}$. The bands show the LHCb measurement (with the inner band showing the statistical uncertainty and the outer band showing the total uncertainty). The points correspond to different theoretical predictions with the error bars indicating their uncertainties as described in the main text. These results are corrected for FSR from the final state muons from the $\PZ$ boson decay.\normalsize}
  \label{ratios}
\end{figure}

\begin{figure}
  \centering   
  \includegraphics[width=0.65\textwidth]{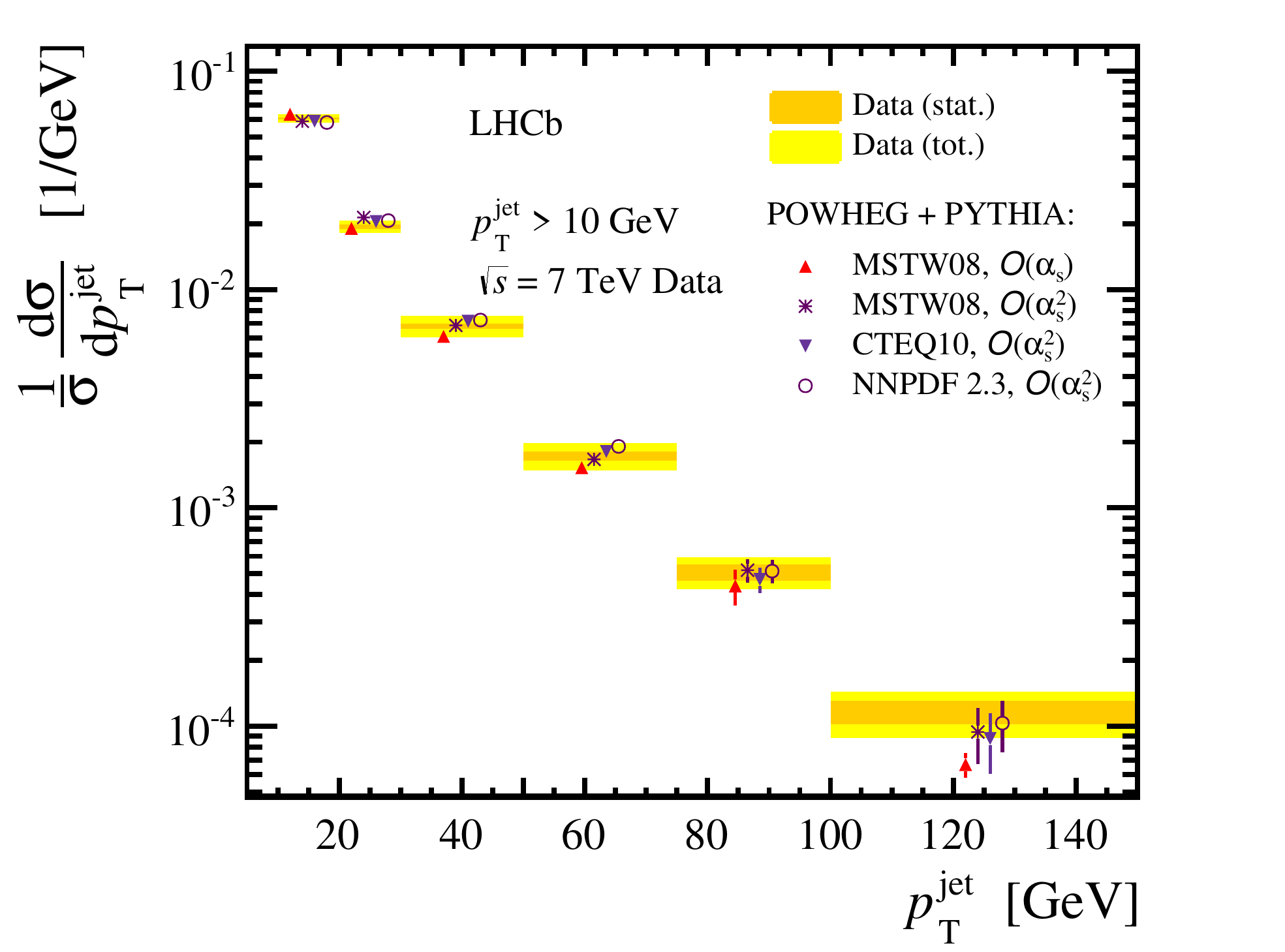}
  \caption{\small Cross-section for Z+jet production, differential in the leading jet $p_\text{T}$, for $p_\text{T}^\text{jet} > 10\text{ GeV}$. The bands show the LHCb measurement (with the inner band showing the statistical uncertainty and the outer band showing the total uncertainty). The points correspond to different theoretical predictions with the error bars indicating their uncertainties as described in the main text. Predictions are displaced horizontally for presentation. These results are not corrected for FSR from the final state muons from the $\PZ$ boson decay.\normalsize}
  \label{res1}
\end{figure}

\begin{figure}
  \centering   
  \includegraphics[width=0.495\textwidth]{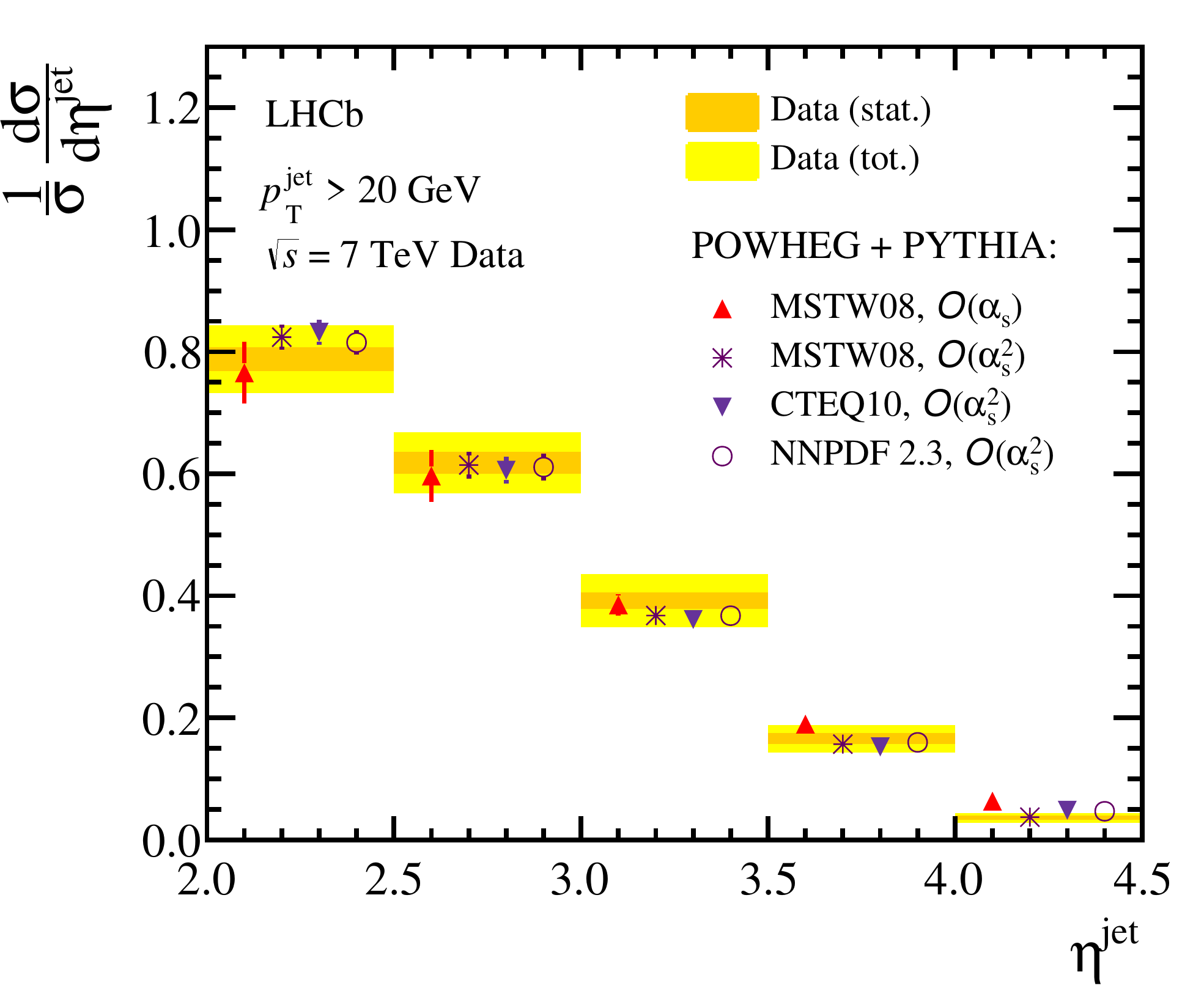}
  \includegraphics[width=0.495\textwidth]{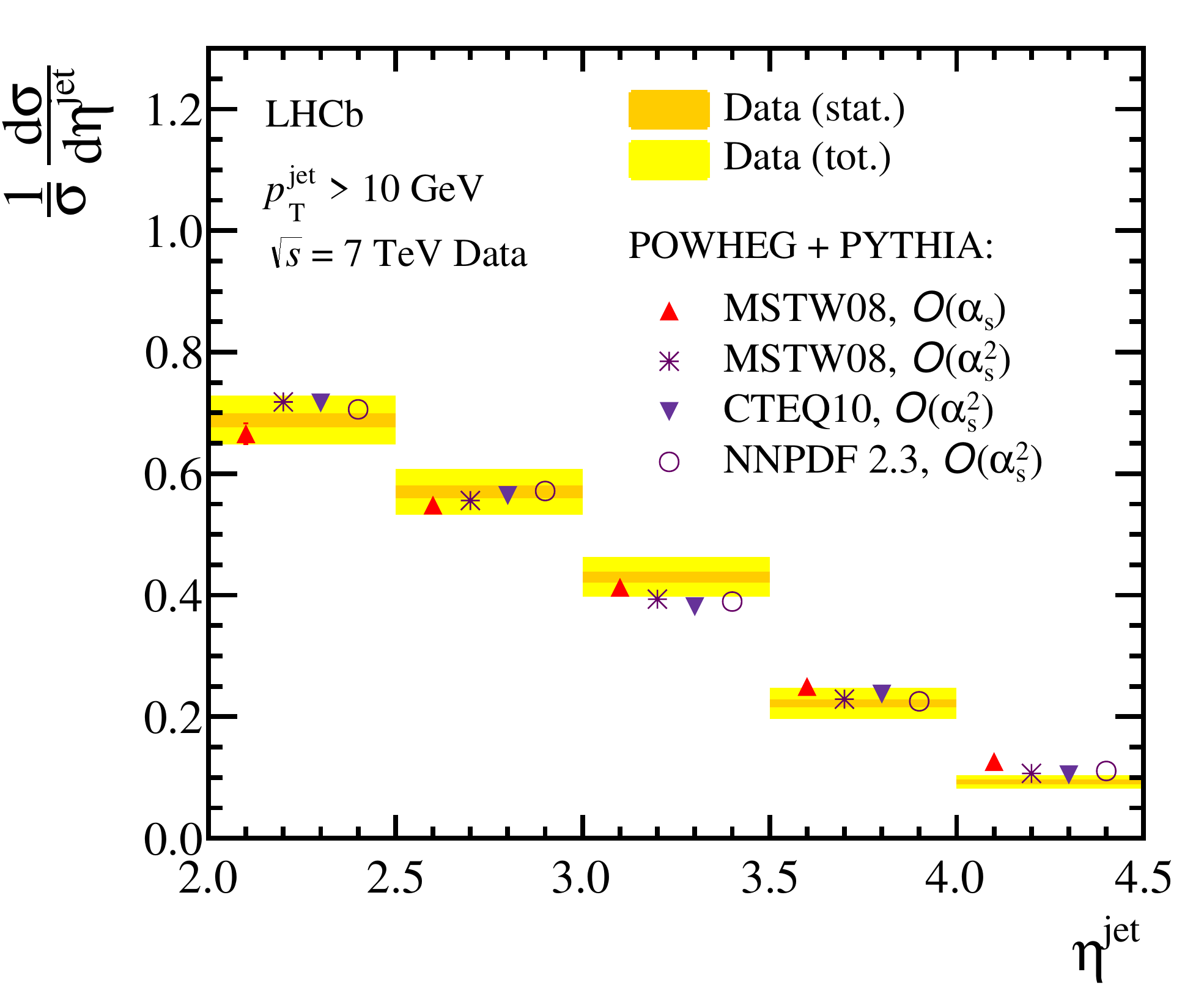}
  \caption{\small Cross-section for Z+jet production, differential in the leading jet pseudorapidity, for (left) $p_\text{T}^\text{jet}\nobreak>\nobreak 20\text{ GeV}$ and (right) $p_\text{T}^\text{jet} > 10\text{ GeV}$. The bands show the LHCb measurement (with the inner band showing the statistical uncertainty and the outer band showing the total uncertainty). Superimposed are predictions as described in Fig.~\ref{res1}.\normalsize}
  \label{res2}
\end{figure}

\begin{figure}
  \centering   
  \includegraphics[width=0.495\textwidth]{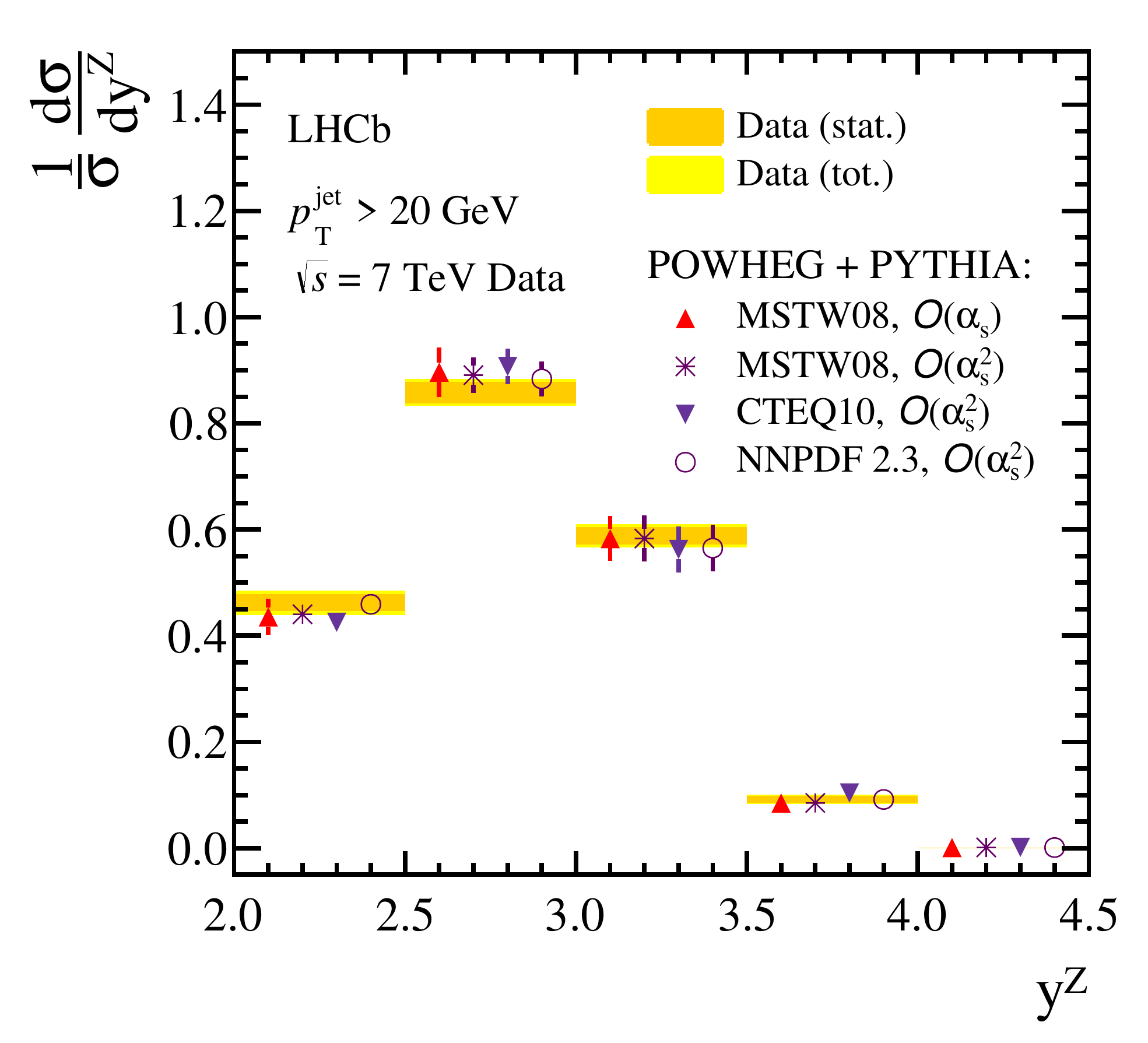}
  \includegraphics[width=0.495\textwidth]{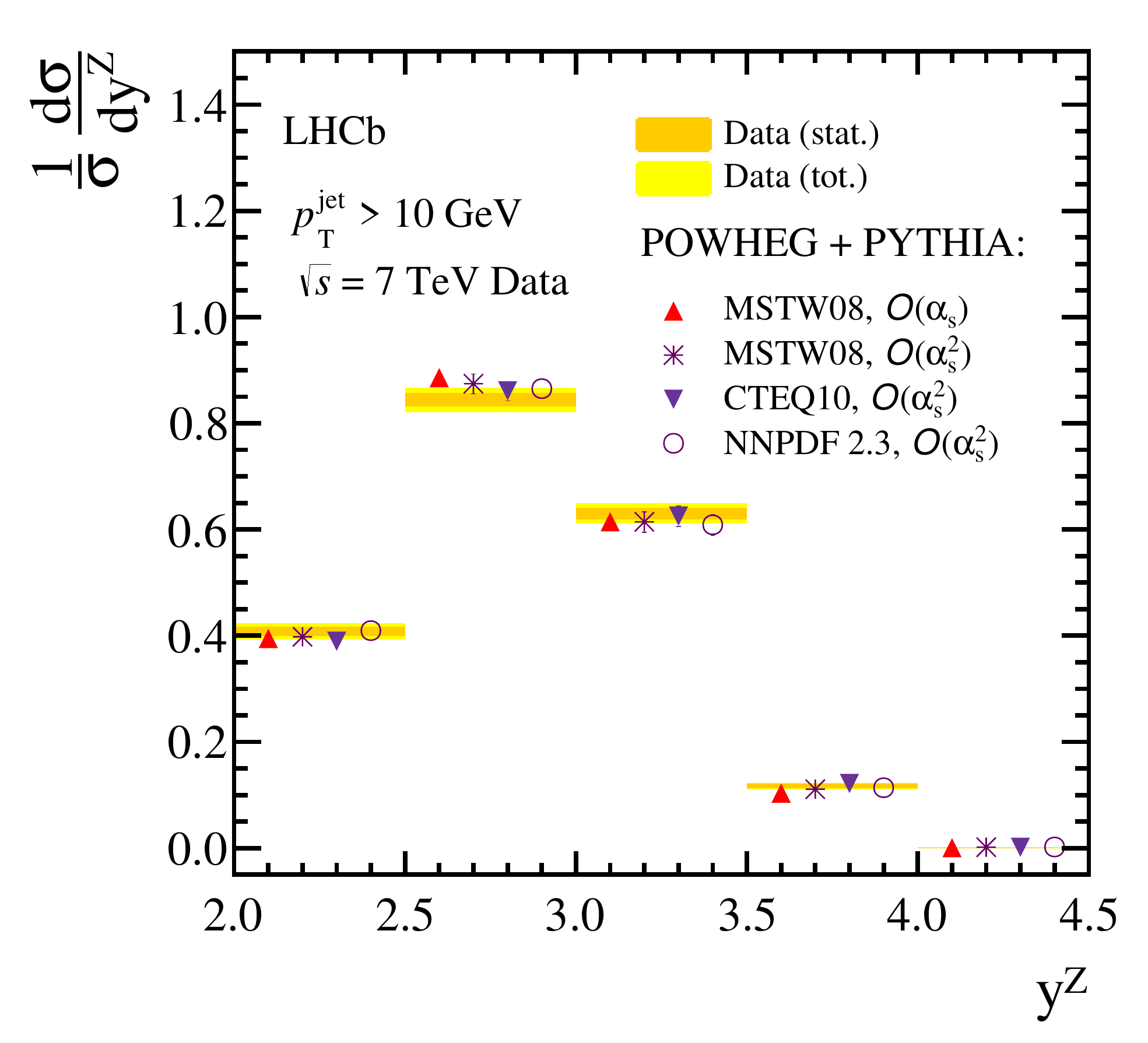}
  \caption{\small Cross-section for Z+jet production, differential in the $\PZ$ boson rapidity, $y^\PZ$, for (left) $p_\text{T}^\text{jet}\nobreak>\nobreak 20\text{ GeV}$ and (right) $p_\text{T}^\text{jet} > 10\text{ GeV}$. The bands show the LHCb measurement (with the inner band showing the statistical uncertainty and the outer band showing the total uncertainty). Superimposed are predictions as described in Fig.~\ref{res1}.\normalsize}
  \label{res3}
\end{figure}

\begin{figure}
  \centering   
  \includegraphics[width=0.495\textwidth]{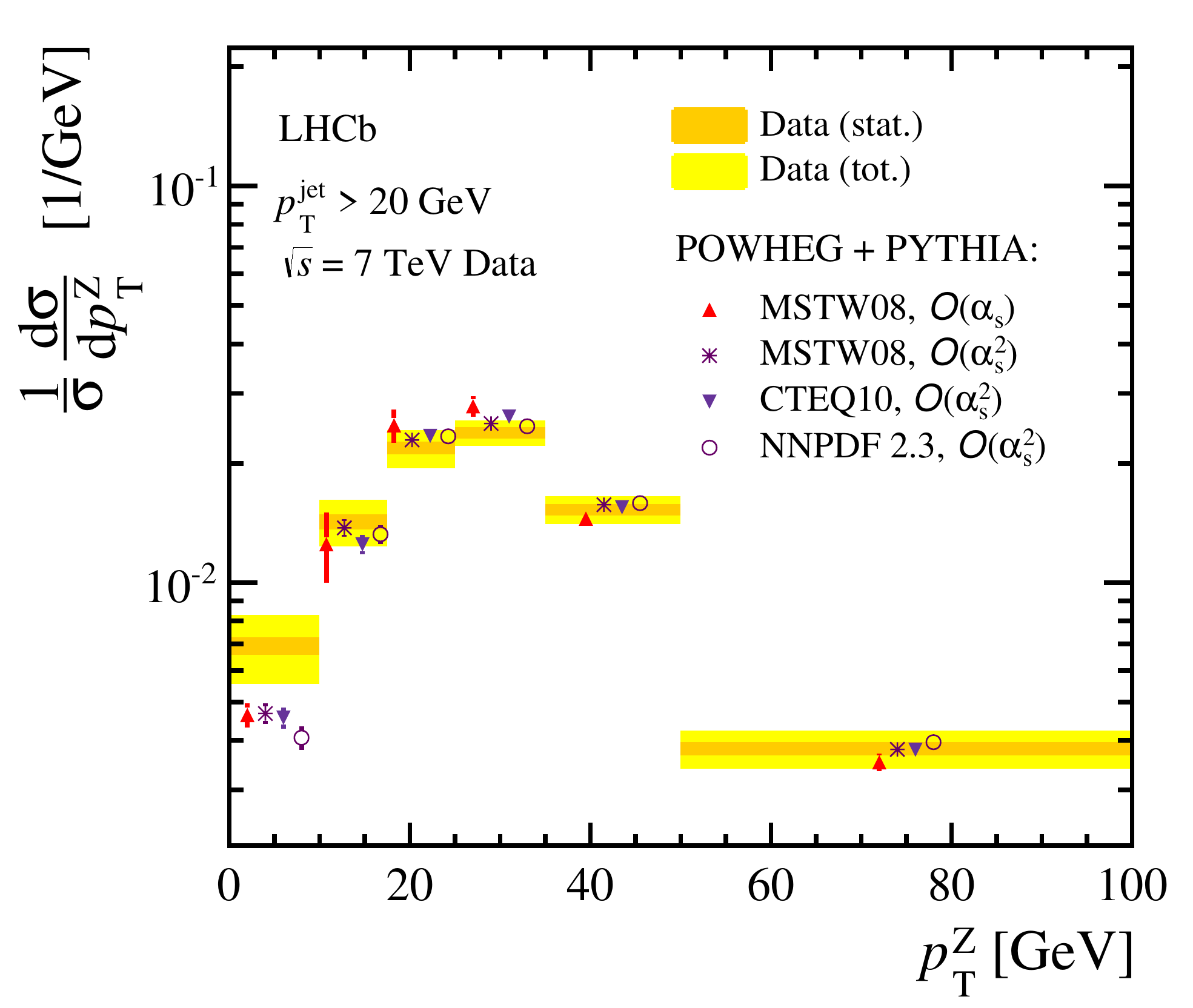}
  \includegraphics[width=0.495\textwidth]{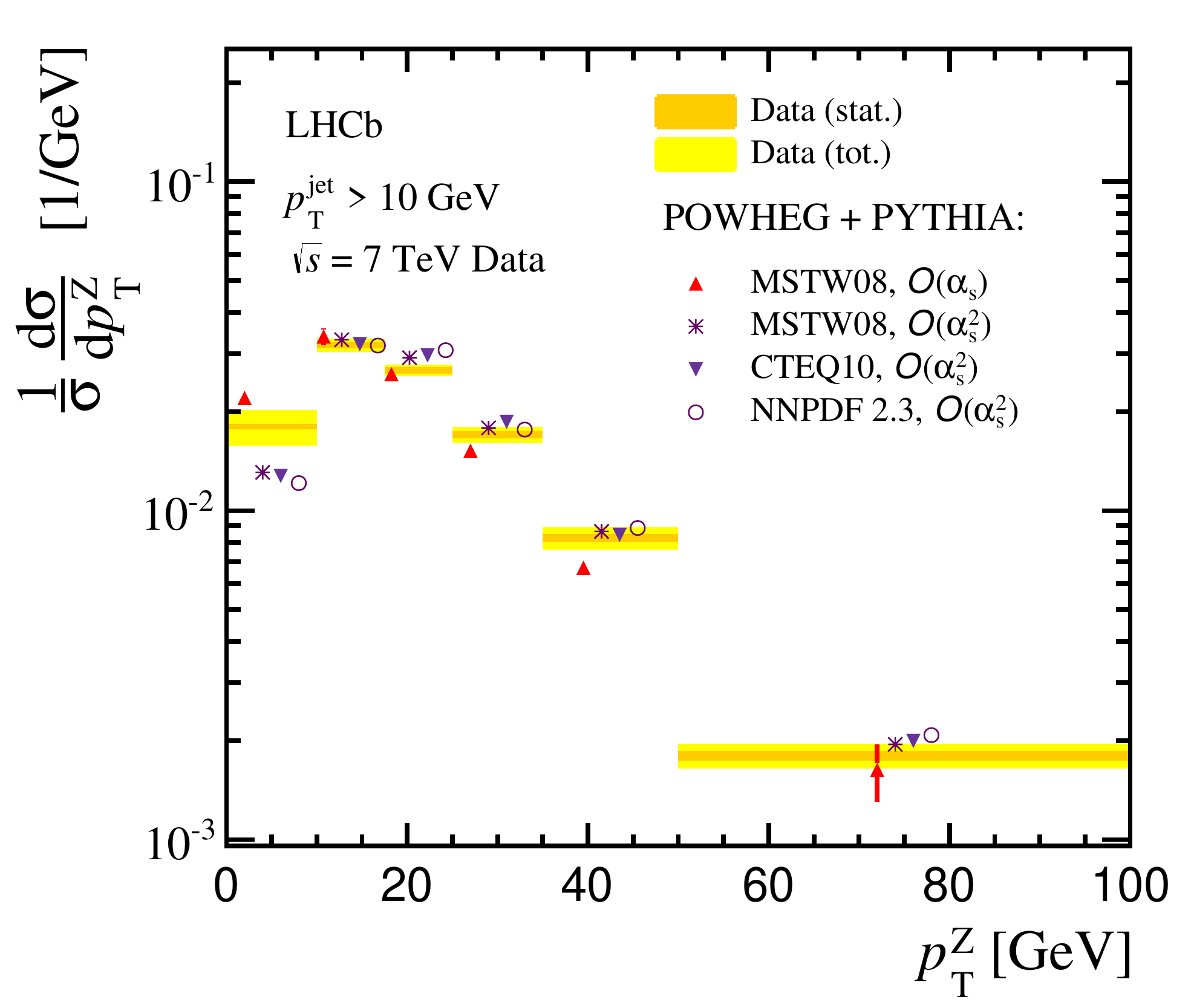}
  \caption{\small Cross-section for Z+jet production, differential in the $\PZ$ boson transverse momentum, for (left) $p_\text{T}^\text{jet}\nobreak>\nobreak 20\text{ GeV}$ and (right) $p_\text{T}^\text{jet} > 10\text{ GeV}$. The bands show the LHCb measurement (with the inner band showing the statistical uncertainty and the outer band showing the total uncertainty). Superimposed are predictions as described in Fig.~\ref{res1}.\normalsize}
  \label{res4}
\end{figure}

\begin{figure}
  \centering   
  \includegraphics[width=0.495\textwidth]{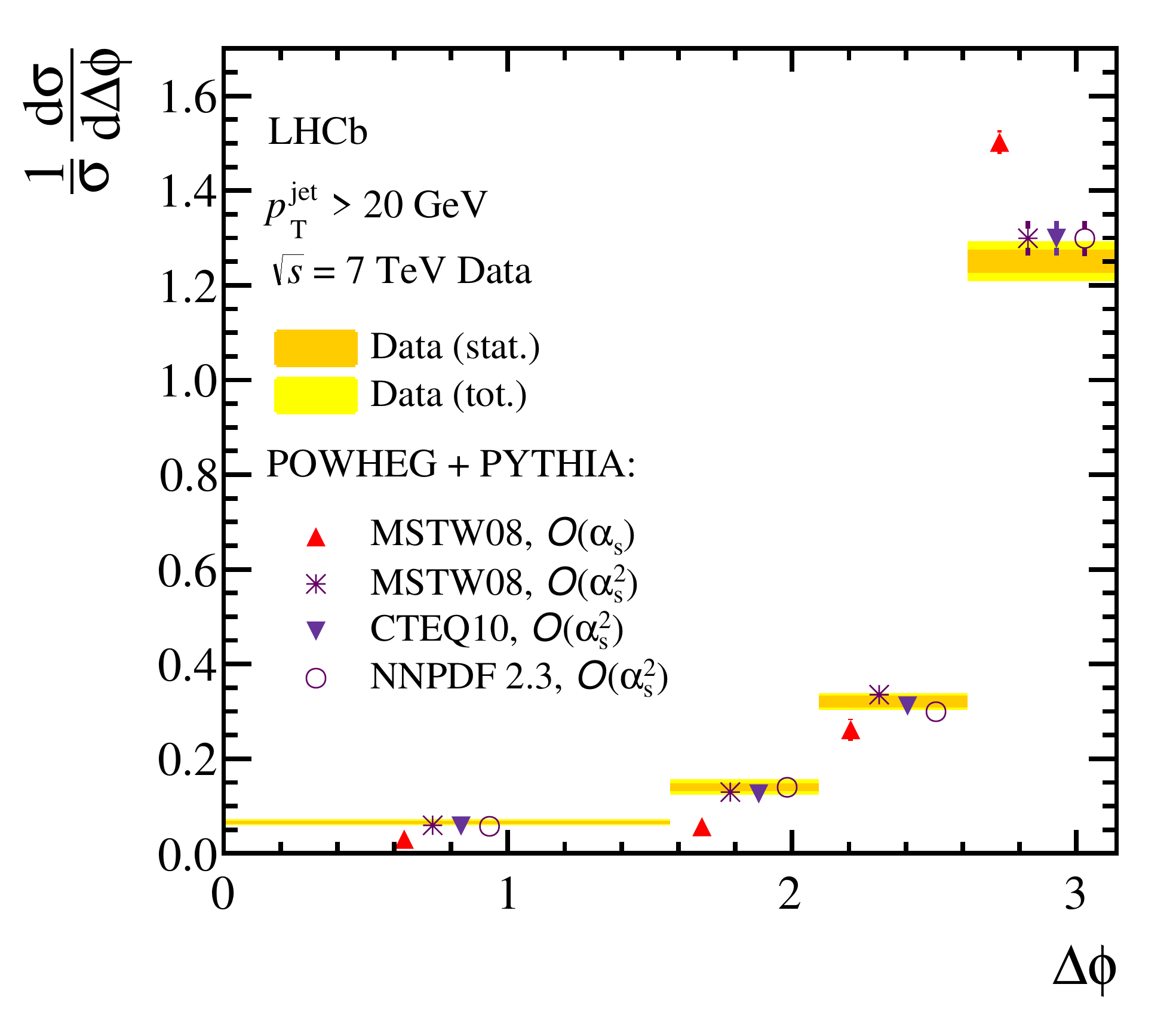}
  \includegraphics[width=0.495\textwidth]{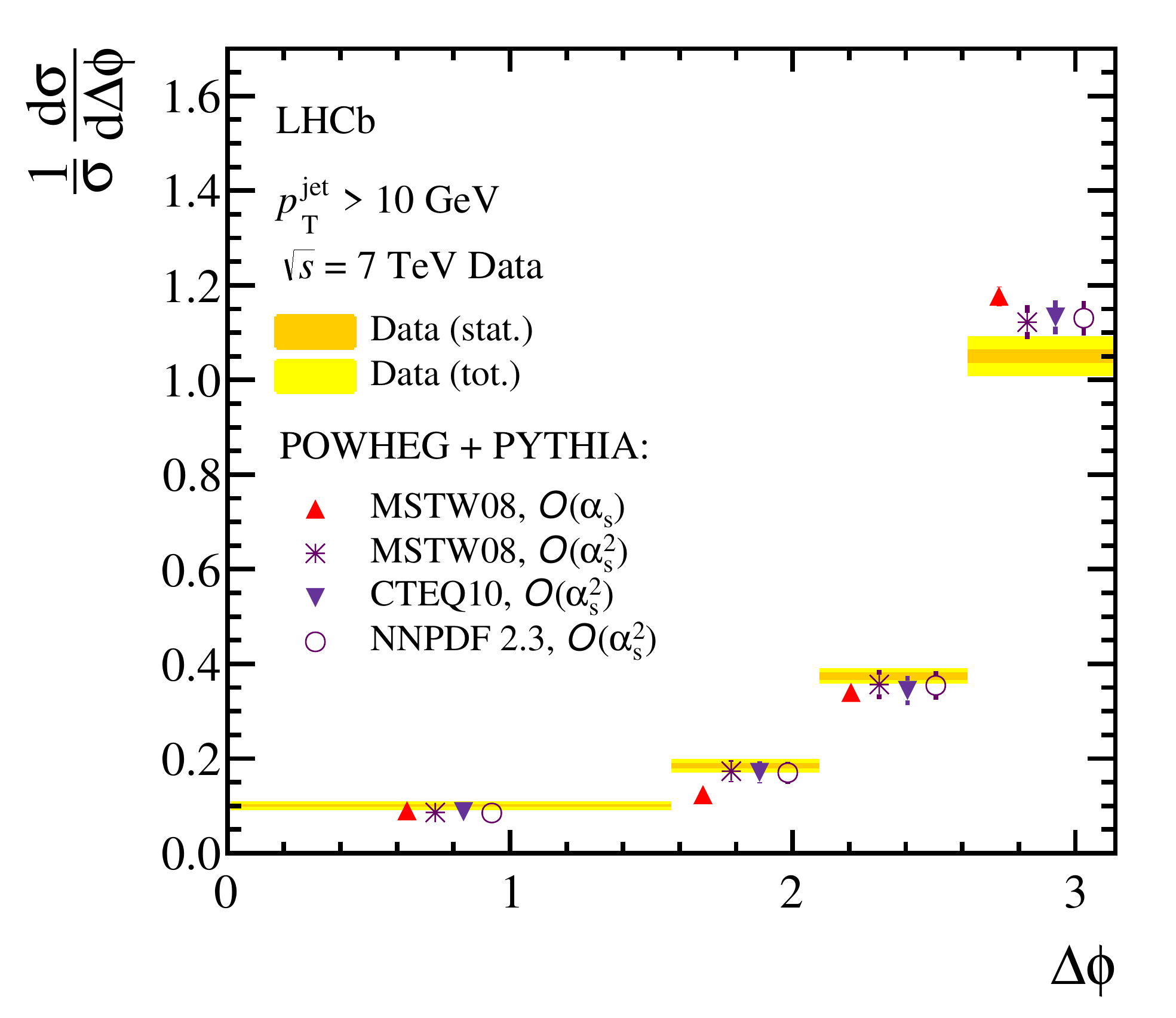}
  \caption{\small Cross-section for Z+jet production, differential in the difference in $\phi$ between the $\PZ$ boson and the leading jet, for (left) $p_\text{T}^\text{jet}\nobreak>\nobreak 20\text{ GeV}$ and (right) $p_\text{T}^\text{jet} > 10\text{ GeV}$. The bands show the LHCb measurement (with the inner band showing the statistical uncertainty and the outer band showing the total uncertainty). Superimposed are predictions as described in Fig.~\ref{res1}.\normalsize}
  \label{res5}
\end{figure}

\begin{figure}
  \centering   
  \includegraphics[width=0.495\textwidth]{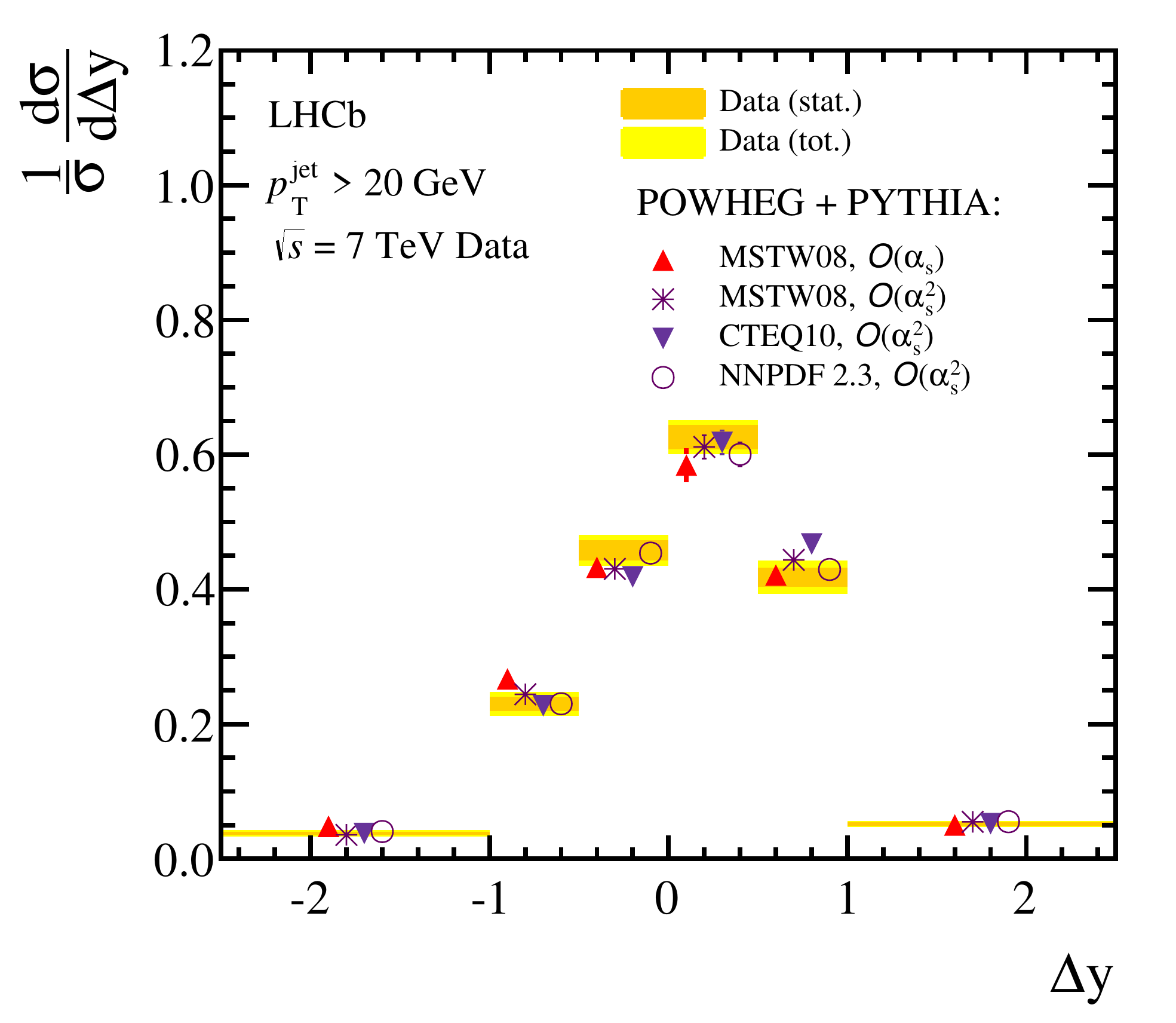}
  \includegraphics[width=0.495\textwidth]{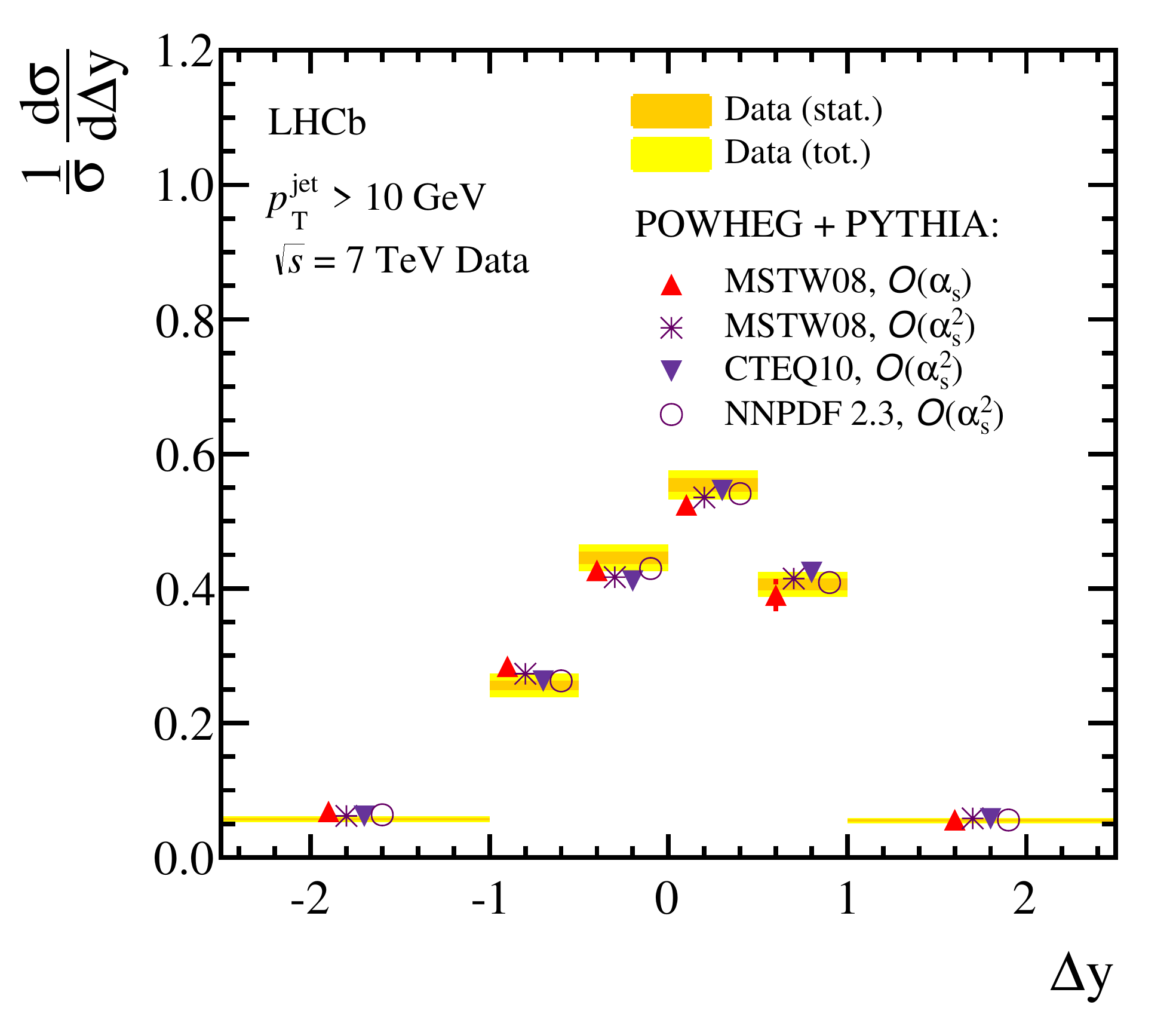}
  \caption{\small Cross-section for Z+jet production, differential in the difference in rapidity between the $\PZ$ boson and the leading jet, for (left) $p_\text{T}^\text{jet}\nobreak>\nobreak 20\text{ GeV}$ and (right) $p_\text{T}^\text{jet} > 10\text{ GeV}$. The bands show the LHCb measurement (with the inner band showing the statistical uncertainty and the outer band showing the total uncertainty).  Superimposed are predictions as described in Fig.~\ref{res1}.\normalsize}
  \label{res6}
\end{figure}

The cross-section ratios are compared in Fig.~\ref{ratios} to the Standard Model theoretical predictions discussed above. The results for the differential cross-sections, uncorrected for final state radiation from the muons, are presented in Figs.~\ref{res1}--\ref{res6}. For all cases reasonable agreement is seen between the Standard Model calculations and the data. The $\mathcal{O}(\alpha_s^2)$ predictions tend to give better agreement with data than the $\mathcal{O}(\alpha_s)$ prediction. This is most noticeably seen in the $\PZ$ boson transverse momentum distribution, shown in Fig.~\ref{res4}. For high values of the boson transverse momentum, the $\mathcal{O}(\alpha_s^2)$ predictions have a slope compatible with that in data, whereas the $\mathcal{O}(\alpha_s)$ prediction is steeper than data. The $\mathcal{O}(\alpha_s^2)$ predictions also match the data better for the $\Delta\phi$ distribution, as shown in Fig.~\ref{res5}. The $\mathcal{O}(\alpha_s)$ prediction overestimates the number of events where the $\PZ$ boson and jet are azimuthally opposed. Higher orders in pQCD are needed to simulate the production of $\PZ$ bosons and jets that are not produced back-to-back as the parton shower tends to produce partons collinear with the parton produced in the hard interaction. Whilst the different PDF parametrisations studied agree with the data, there are hints of tension between the PDF sets in the $\Delta y$ distribution, shown in Fig.~\ref{res6}.

\section{Summary}
\label{sec:Conclusion}

A measurement of the $\Pp\Pp\rightarrow \PZ(\rightarrow\mup\mun)+\text{jet}$ production cross-section at $\sqrt{s} = 7$~TeV is presented, using a data sample corresponding to an integrated luminosity of $1.0\,$fb$^{-1}$ recorded by the LHCb experiment. The measurement is performed within the kinematic acceptance, $p_\text{T}^\mu > 20$~GeV, $2.0 < \eta^\mu < 4.5$, $60<M_{\mu\mu}<120$~GeV, $2.0<\eta^\text{jet}<4.5$ and $\Delta r(\mu,\text{jet})\nolinebreak >\nolinebreak 0.4$. The cross-sections are determined for jets with transverse momenta exceeding two thresholds, 20 and 10~GeV. The differential cross-sections are also measured as a function of various variables describing the Z boson kinematic properties, the jet kinematic properties, and the correlations between them. The measured cross-sections show reasonable agreement with expectations from $\mathcal{O}(\alpha_s^2)$ calculations, for all the PDF parametrisations studied. Predictions at $\mathcal{O}(\alpha_s^2)$ show better agreement with the $p_\text{T}$ and $\Delta\phi$ distributions, which are sensitive to higher order effects, than predictions at $\mathcal{O}(\alpha_s)$.
\clearpage

\section*{Acknowledgements}

\noindent We express our gratitude to our colleagues in the CERN
accelerator departments for the excellent performance of the LHC. We
thank the technical and administrative staff at the LHCb
institutes. We acknowledge support from CERN and from the national
agencies: CAPES, CNPq, FAPERJ and FINEP (Brazil); NSFC (China);
CNRS/IN2P3 and Region Auvergne (France); BMBF, DFG, HGF and MPG
(Germany); SFI (Ireland); INFN (Italy); FOM and NWO (The Netherlands);
SCSR (Poland); MEN/IFA (Romania); MinES, Rosatom, RFBR and NRC
``Kurchatov Institute'' (Russia); MinECo, XuntaGal and GENCAT (Spain);
SNSF and SER (Switzerland); NAS Ukraine (Ukraine); STFC (United
Kingdom); NSF (USA). We also acknowledge the support received from the
ERC under FP7. The Tier1 computing centres are supported by IN2P3
(France), KIT and BMBF (Germany), INFN (Italy), NWO and SURF (The
Netherlands), PIC (Spain), GridPP (United Kingdom). We are thankful
for the computing resources put at our disposal by Yandex LLC
(Russia), as well as to the communities behind the multiple open
source software packages that we depend on.

\addcontentsline{toc}{section}{References}
\setboolean{inbibliography}{true}
\bibliographystyle{LHCb}
\bibliography{main,LHCb-PAPER,LHCb-CONF,LHCb-DP}

\ifx\mcitethebibliography\mciteundefinedmacro
\PackageError{LHCb.bst}{mciteplus.sty has not been loaded}
{This bibstyle requires the use of the mciteplus package.}\fi
\providecommand{\href}[2]{#2}
\begin{mcitethebibliography}{10}
\mciteSetBstSublistMode{n}
\mciteSetBstMaxWidthForm{subitem}{\alph{mcitesubitemcount})}
\mciteSetBstSublistLabelBeginEnd{\mcitemaxwidthsubitemform\space}
{\relax}{\relax}

\bibitem{Thorne:2008am}
R.~Thorne, A.~Martin, W.~Stirling, and G.~Watt,
  \ifthenelse{\boolean{articletitles}}{{\it {Parton distributions and QCD at
  LHCb}}, }{} in {\em Proc.~of XVI Int.~Workshop on Deep-Inelastic Scattering
  and Related Topics}, p.~30, 2008.
\newblock \href{http://arxiv.org/abs/0808.1847}{{\tt arXiv:0808.1847}}.
\newblock
  doi:~\href{http://dx.doi.org/10.3360/dis.2008.30}{10.3360/dis.2008.30}\relax
\mciteBstWouldAddEndPuncttrue
\mciteSetBstMidEndSepPunct{\mcitedefaultmidpunct}
{\mcitedefaultendpunct}{\mcitedefaultseppunct}\relax
\EndOfBibitem
\bibitem{LHCb-PAPER-2012-008}
LHCb collaboration, R.~Aaij {\em et~al.},
  \ifthenelse{\boolean{articletitles}}{{\it {Inclusive $W$ and $Z$ production
  in the forward region at $\sqrt{s} = 7\tev$}},
  }{}\href{http://dx.doi.org/10.1007/JHEP06(2012)058}{JHEP {\bf 06} (2012) 58},
  \href{http://arxiv.org/abs/1204.1620}{{\tt arXiv:1204.1620}}\relax
\mciteBstWouldAddEndPuncttrue
\mciteSetBstMidEndSepPunct{\mcitedefaultmidpunct}
{\mcitedefaultendpunct}{\mcitedefaultseppunct}\relax
\EndOfBibitem
\bibitem{LHCb-PAPER-2012-036}
LHCb collaboration, R.~Aaij {\em et~al.},
  \ifthenelse{\boolean{articletitles}}{{\it {Measurement of the cross-section
  for $Z \to e^+e^-$ production in $pp$ collisions at $\sqrt{s} = 7 \tev$}},
  }{}\href{http://dx.doi.org/10.1007/JHEP02(2013)106}{JHEP {\bf 02} (2013)
  106}, \href{http://arxiv.org/abs/1212.4620}{{\tt arXiv:1212.4620}}\relax
\mciteBstWouldAddEndPuncttrue
\mciteSetBstMidEndSepPunct{\mcitedefaultmidpunct}
{\mcitedefaultendpunct}{\mcitedefaultseppunct}\relax
\EndOfBibitem
\bibitem{LHCb-PAPER-2012-029}
LHCb collaboration, R.~Aaij {\em et~al.},
  \ifthenelse{\boolean{articletitles}}{{\it {A study of the $Z$ production
  cross-section in $pp$ collisions at $\sqrt{s}=7 \tev$ using tau final
  states}}, }{}\href{http://dx.doi.org/10.1007/JHEP01(2013)111}{JHEP {\bf 01}
  (2013) 111}, \href{http://arxiv.org/abs/1210.6289}{{\tt
  arXiv:1210.6289}}\relax
\mciteBstWouldAddEndPuncttrue
\mciteSetBstMidEndSepPunct{\mcitedefaultmidpunct}
{\mcitedefaultendpunct}{\mcitedefaultseppunct}\relax
\EndOfBibitem
\bibitem{JuanTalk}
S.~A. Malik and G.~Watt, \ifthenelse{\boolean{articletitles}}{{\it {Ratios of
  $W$ and $Z$ cross sections at large boson $p_T$ as a constraint on PDFs and
  background to new physics}}, }{}\href{http://arxiv.org/abs/1304.2424}{{\tt
  arXiv:1304.2424}}\relax
\mciteBstWouldAddEndPuncttrue
\mciteSetBstMidEndSepPunct{\mcitedefaultmidpunct}
{\mcitedefaultendpunct}{\mcitedefaultseppunct}\relax
\EndOfBibitem
\bibitem{Aad:2013ysa}
ATLAS collaboration, G.~Aad {\em et~al.},
  \ifthenelse{\boolean{articletitles}}{{\it {Measurement of the production
  cross section of jets in association with a Z boson in pp collisions at
  $\sqrt{s} = 7$ TeV with the ATLAS detector}},
  }{}\href{http://dx.doi.org/10.1007/JHEP07(2013)032}{JHEP {\bf 07} (2013)
  032}, \href{http://arxiv.org/abs/1304.7098}{{\tt arXiv:1304.7098}}\relax
\mciteBstWouldAddEndPuncttrue
\mciteSetBstMidEndSepPunct{\mcitedefaultmidpunct}
{\mcitedefaultendpunct}{\mcitedefaultseppunct}\relax
\EndOfBibitem
\bibitem{Chatrchyan:2013tna}
CMS Collaboration, S.~Chatrchyan {\em et~al.},
  \ifthenelse{\boolean{articletitles}}{{\it {Event shapes and azimuthal
  correlations in $Z$ + jets events in $pp$ collisions at $\sqrt{s}=7$ TeV}},
  }{}\href{http://dx.doi.org/10.1016/j.physletb.2013.04.025}{Phys.\ Lett.\
  {\bf B722} (2013) 238}, \href{http://arxiv.org/abs/1301.1646}{{\tt
  arXiv:1301.1646}}\relax
\mciteBstWouldAddEndPuncttrue
\mciteSetBstMidEndSepPunct{\mcitedefaultmidpunct}
{\mcitedefaultendpunct}{\mcitedefaultseppunct}\relax
\EndOfBibitem
\bibitem{Chatrchyan:2013oda}
CMS Collaboration, S.~Chatrchyan {\em et~al.},
  \ifthenelse{\boolean{articletitles}}{{\it {Rapidity distributions in
  exclusive Z + jet and photon + jet events in pp collisions at sqrt(s)=7
  TeV}}, }{}\href{http://arxiv.org/abs/1310.3082}{{\tt arXiv:1310.3082}}\relax
\mciteBstWouldAddEndPuncttrue
\mciteSetBstMidEndSepPunct{\mcitedefaultmidpunct}
{\mcitedefaultendpunct}{\mcitedefaultseppunct}\relax
\EndOfBibitem
\bibitem{Hautmann:2012sh}
F.~Hautmann, M.~Hentschinski, and H.~Jung,
  \ifthenelse{\boolean{articletitles}}{{\it {Forward Z-boson production and the
  unintegrated sea quark density}},
  }{}\href{http://dx.doi.org/10.1016/j.nuclphysb.2012.07.023}{Nucl.\ Phys.\
  {\bf B865} (2012) 54}, \href{http://arxiv.org/abs/1205.1759}{{\tt
  arXiv:1205.1759}}\relax
\mciteBstWouldAddEndPuncttrue
\mciteSetBstMidEndSepPunct{\mcitedefaultmidpunct}
{\mcitedefaultendpunct}{\mcitedefaultseppunct}\relax
\EndOfBibitem
\bibitem{Arnold:1988dp}
P.~B. Arnold and M.~H. Reno, \ifthenelse{\boolean{articletitles}}{{\it {The
  complete computation of high $p_\text{t}$ W and Z production in second-order
  QCD}}, }{}\href{http://dx.doi.org/10.1016/0550-3213(89)90600-7}{Nucl.\ Phys.\
   {\bf B319} (1989) 37}\relax
\mciteBstWouldAddEndPuncttrue
\mciteSetBstMidEndSepPunct{\mcitedefaultmidpunct}
{\mcitedefaultendpunct}{\mcitedefaultseppunct}\relax
\EndOfBibitem
\bibitem{Giele:1993dj}
W.~Giele, E.~N. Glover, and D.~A. Kosower,
  \ifthenelse{\boolean{articletitles}}{{\it {Higher order corrections to jet
  cross-sections in hadron colliders}},
  }{}\href{http://dx.doi.org/10.1016/0550-3213(93)90365-V}{Nucl.\ Phys.\  {\bf
  B403} (1993) 633}, \href{http://arxiv.org/abs/hep-ph/9302225}{{\tt
  arXiv:hep-ph/9302225}}\relax
\mciteBstWouldAddEndPuncttrue
\mciteSetBstMidEndSepPunct{\mcitedefaultmidpunct}
{\mcitedefaultendpunct}{\mcitedefaultseppunct}\relax
\EndOfBibitem
\bibitem{Anastasiou:2003ds}
C.~Anastasiou, L.~J. Dixon, K.~Melnikov, and F.~Petriello,
  \ifthenelse{\boolean{articletitles}}{{\it {High precision QCD at hadron
  colliders: electroweak gauge boson rapidity distributions at NNLO}},
  }{}\href{http://dx.doi.org/10.1103/PhysRevD.69.094008}{Phys.\ Rev.\  {\bf
  D69} (2004) 094008}, \href{http://arxiv.org/abs/hep-ph/0312266}{{\tt
  arXiv:hep-ph/0312266}}\relax
\mciteBstWouldAddEndPuncttrue
\mciteSetBstMidEndSepPunct{\mcitedefaultmidpunct}
{\mcitedefaultendpunct}{\mcitedefaultseppunct}\relax
\EndOfBibitem
\bibitem{Gavin:2010az}
R.~Gavin, Y.~Li, F.~Petriello, and S.~Quackenbush,
  \ifthenelse{\boolean{articletitles}}{{\it {FEWZ 2.0: A code for hadronic Z
  production at next-to-next-to-leading order}},
  }{}\href{http://dx.doi.org/10.1016/j.cpc.2011.06.008}{Comput.\ Phys.\
  Commun.\  {\bf 182} (2011) 2388}, \href{http://arxiv.org/abs/1011.3540}{{\tt
  arXiv:1011.3540}}\relax
\mciteBstWouldAddEndPuncttrue
\mciteSetBstMidEndSepPunct{\mcitedefaultmidpunct}
{\mcitedefaultendpunct}{\mcitedefaultseppunct}\relax
\EndOfBibitem
\bibitem{Campbell:2010ff}
J.~M. Campbell and R.~Ellis, \ifthenelse{\boolean{articletitles}}{{\it {MCFM
  for the Tevatron and the LHC}},
  }{}\href{http://dx.doi.org/10.1016/j.nuclphysbps.2010.08.011}{Nucl.\ Phys.\
  Proc.\ Suppl.\  {\bf 205-206} (2010) 10},
  \href{http://arxiv.org/abs/1007.3492}{{\tt arXiv:1007.3492}}\relax
\mciteBstWouldAddEndPuncttrue
\mciteSetBstMidEndSepPunct{\mcitedefaultmidpunct}
{\mcitedefaultendpunct}{\mcitedefaultseppunct}\relax
\EndOfBibitem
\bibitem{PowhegZj}
S.~Alioli, P.~Nason, C.~Oleari, and E.~Re,
  \ifthenelse{\boolean{articletitles}}{{\it {Vector boson plus one jet
  production in POWHEG}},
  }{}\href{http://dx.doi.org/10.1007/JHEP01(2011)095}{JHEP {\bf 01} (2011)
  095}, \href{http://arxiv.org/abs/1009.5594}{{\tt arXiv:1009.5594}}\relax
\mciteBstWouldAddEndPuncttrue
\mciteSetBstMidEndSepPunct{\mcitedefaultmidpunct}
{\mcitedefaultendpunct}{\mcitedefaultseppunct}\relax
\EndOfBibitem
\bibitem{Mangano:2002ea}
M.~L. Mangano {\em et~al.}, \ifthenelse{\boolean{articletitles}}{{\it {ALPGEN,
  a generator for hard multiparton processes in hadronic collisions}}, }{}JHEP
  {\bf 07} (2003) 001, \href{http://arxiv.org/abs/hep-ph/0206293}{{\tt
  arXiv:hep-ph/0206293}}\relax
\mciteBstWouldAddEndPuncttrue
\mciteSetBstMidEndSepPunct{\mcitedefaultmidpunct}
{\mcitedefaultendpunct}{\mcitedefaultseppunct}\relax
\EndOfBibitem
\bibitem{Gleisberg:2008ta}
T.~Gleisberg {\em et~al.}, \ifthenelse{\boolean{articletitles}}{{\it {Event
  generation with SHERPA 1.1}},
  }{}\href{http://dx.doi.org/10.1088/1126-6708/2009/02/007}{JHEP {\bf 02}
  (2009) 007}, \href{http://arxiv.org/abs/0811.4622}{{\tt
  arXiv:0811.4622}}\relax
\mciteBstWouldAddEndPuncttrue
\mciteSetBstMidEndSepPunct{\mcitedefaultmidpunct}
{\mcitedefaultendpunct}{\mcitedefaultseppunct}\relax
\EndOfBibitem
\bibitem{Cacciari:2008gp}
M.~Cacciari, G.~P. Salam, and G.~Soyez,
  \ifthenelse{\boolean{articletitles}}{{\it {The anti-$k_t$ jet clustering
  algorithm}}, }{}\href{http://dx.doi.org/10.1088/1126-6708/2008/04/063}{JHEP
  {\bf 04} (2008) 063}, \href{http://arxiv.org/abs/0802.1189}{{\tt
  arXiv:0802.1189}}\relax
\mciteBstWouldAddEndPuncttrue
\mciteSetBstMidEndSepPunct{\mcitedefaultmidpunct}
{\mcitedefaultendpunct}{\mcitedefaultseppunct}\relax
\EndOfBibitem
\bibitem{Alves:2008zz}
LHCb collaboration, A.~A. Alves~Jr. {\em et~al.},
  \ifthenelse{\boolean{articletitles}}{{\it {The \lhcb detector at the LHC}},
  }{}\href{http://dx.doi.org/10.1088/1748-0221/3/08/S08005}{JINST {\bf 3}
  (2008) S08005}\relax
\mciteBstWouldAddEndPuncttrue
\mciteSetBstMidEndSepPunct{\mcitedefaultmidpunct}
{\mcitedefaultendpunct}{\mcitedefaultseppunct}\relax
\EndOfBibitem
\bibitem{LHCb-DP-2012-003}
M.~Adinolfi {\em et~al.}, \ifthenelse{\boolean{articletitles}}{{\it
  {Performance of the \lhcb RICH detector at the LHC}},
  }{}\href{http://dx.doi.org/10.1140/epjc/s10052-013-2431-9}{Eur.\ Phys.\ J.\
  {\bf C73} (2013) 2431}, \href{http://arxiv.org/abs/1211.6759}{{\tt
  arXiv:1211.6759}}\relax
\mciteBstWouldAddEndPuncttrue
\mciteSetBstMidEndSepPunct{\mcitedefaultmidpunct}
{\mcitedefaultendpunct}{\mcitedefaultseppunct}\relax
\EndOfBibitem
\bibitem{LHCb-DP-2012-002}
A.~A. Alves~Jr {\em et~al.}, \ifthenelse{\boolean{articletitles}}{{\it
  {Performance of the LHCb muon system}},
  }{}\href{http://dx.doi.org/10.1088/1748-0221/8/02/P02022}{JINST {\bf 8}
  (2013) P02022}, \href{http://arxiv.org/abs/1211.1346}{{\tt
  arXiv:1211.1346}}\relax
\mciteBstWouldAddEndPuncttrue
\mciteSetBstMidEndSepPunct{\mcitedefaultmidpunct}
{\mcitedefaultendpunct}{\mcitedefaultseppunct}\relax
\EndOfBibitem
\bibitem{LHCb-DP-2012-004}
R.~Aaij {\em et~al.}, \ifthenelse{\boolean{articletitles}}{{\it {The \lhcb
  trigger and its performance in 2011}},
  }{}\href{http://dx.doi.org/10.1088/1748-0221/8/04/P04022}{JINST {\bf 8}
  (2013) P04022}, \href{http://arxiv.org/abs/1211.3055}{{\tt
  arXiv:1211.3055}}\relax
\mciteBstWouldAddEndPuncttrue
\mciteSetBstMidEndSepPunct{\mcitedefaultmidpunct}
{\mcitedefaultendpunct}{\mcitedefaultseppunct}\relax
\EndOfBibitem
\bibitem{Sjostrand:2006za}
T.~Sj\"{o}strand, S.~Mrenna, and P.~Skands,
  \ifthenelse{\boolean{articletitles}}{{\it {PYTHIA 6.4 physics and manual}},
  }{}\href{http://dx.doi.org/10.1088/1126-6708/2006/05/026}{JHEP {\bf 05}
  (2006) 026}, \href{http://arxiv.org/abs/hep-ph/0603175}{{\tt
  arXiv:hep-ph/0603175}}\relax
\mciteBstWouldAddEndPuncttrue
\mciteSetBstMidEndSepPunct{\mcitedefaultmidpunct}
{\mcitedefaultendpunct}{\mcitedefaultseppunct}\relax
\EndOfBibitem
\bibitem{LHCb-PROC-2010-056}
I.~Belyaev {\em et~al.}, \ifthenelse{\boolean{articletitles}}{{\it {Handling of
  the generation of primary events in \gauss, the \lhcb simulation framework}},
  }{}\href{http://dx.doi.org/10.1109/NSSMIC.2010.5873949}{Nuclear Science
  Symposium Conference Record (NSS/MIC) {\bf IEEE} (2010) 1155}\relax
\mciteBstWouldAddEndPuncttrue
\mciteSetBstMidEndSepPunct{\mcitedefaultmidpunct}
{\mcitedefaultendpunct}{\mcitedefaultseppunct}\relax
\EndOfBibitem
\bibitem{Nadolsky:2008zw}
P.~M. Nadolsky {\em et~al.}, \ifthenelse{\boolean{articletitles}}{{\it
  {Implications of CTEQ global analysis for collider observables}},
  }{}\href{http://dx.doi.org/10.1103/PhysRevD.78.013004}{Phys.\ Rev.\  {\bf
  D78} (2008) 013004}, \href{http://arxiv.org/abs/0802.0007}{{\tt
  arXiv:0802.0007}}\relax
\mciteBstWouldAddEndPuncttrue
\mciteSetBstMidEndSepPunct{\mcitedefaultmidpunct}
{\mcitedefaultendpunct}{\mcitedefaultseppunct}\relax
\EndOfBibitem
\bibitem{Lange:2001uf}
D.~J. Lange, \ifthenelse{\boolean{articletitles}}{{\it {The EvtGen particle
  decay simulation package}},
  }{}\href{http://dx.doi.org/10.1016/S0168-9002(01)00089-4}{Nucl.\ Instrum.\
  Meth.\  {\bf A462} (2001) 152}\relax
\mciteBstWouldAddEndPuncttrue
\mciteSetBstMidEndSepPunct{\mcitedefaultmidpunct}
{\mcitedefaultendpunct}{\mcitedefaultseppunct}\relax
\EndOfBibitem
\bibitem{Golonka:2005pn}
P.~Golonka and Z.~Was, \ifthenelse{\boolean{articletitles}}{{\it {PHOTOS Monte
  Carlo: a precision tool for QED corrections in $Z$ and $W$ decays}},
  }{}\href{http://dx.doi.org/10.1140/epjc/s2005-02396-4}{Eur.\ Phys.\ J.\  {\bf
  C45} (2006) 97}, \href{http://arxiv.org/abs/hep-ph/0506026}{{\tt
  arXiv:hep-ph/0506026}}\relax
\mciteBstWouldAddEndPuncttrue
\mciteSetBstMidEndSepPunct{\mcitedefaultmidpunct}
{\mcitedefaultendpunct}{\mcitedefaultseppunct}\relax
\EndOfBibitem
\bibitem{Allison:2006ve}
Geant4 collaboration, J.~Allison {\em et~al.},
  \ifthenelse{\boolean{articletitles}}{{\it {Geant4 developments and
  applications}}, }{}\href{http://dx.doi.org/10.1109/TNS.2006.869826}{IEEE
  Trans.\ Nucl.\ Sci.\  {\bf 53} (2006) 270}\relax
\mciteBstWouldAddEndPuncttrue
\mciteSetBstMidEndSepPunct{\mcitedefaultmidpunct}
{\mcitedefaultendpunct}{\mcitedefaultseppunct}\relax
\EndOfBibitem
\bibitem{Agostinelli:2002hh}
Geant4 collaboration, S.~Agostinelli {\em et~al.},
  \ifthenelse{\boolean{articletitles}}{{\it {Geant4: a simulation toolkit}},
  }{}\href{http://dx.doi.org/10.1016/S0168-9002(03)01368-8}{Nucl.\ Instrum.\
  Meth.\  {\bf A506} (2003) 250}\relax
\mciteBstWouldAddEndPuncttrue
\mciteSetBstMidEndSepPunct{\mcitedefaultmidpunct}
{\mcitedefaultendpunct}{\mcitedefaultseppunct}\relax
\EndOfBibitem
\bibitem{LHCb-PROC-2011-006}
M.~Clemencic {\em et~al.}, \ifthenelse{\boolean{articletitles}}{{\it {The \lhcb
  simulation application, \gauss: design, evolution and experience}},
  }{}\href{http://dx.doi.org/10.1088/1742-6596/331/3/032023}{{J.\ Phys.\ Conf.\
  Ser.\ } {\bf 331} (2011) 032023}\relax
\mciteBstWouldAddEndPuncttrue
\mciteSetBstMidEndSepPunct{\mcitedefaultmidpunct}
{\mcitedefaultendpunct}{\mcitedefaultseppunct}\relax
\EndOfBibitem
\bibitem{Cacciari:2005hq}
M.~Cacciari and G.~P. Salam, \ifthenelse{\boolean{articletitles}}{{\it
  {Dispelling the $N^{3}$ myth for the $k_t$ jet-finder}},
  }{}\href{http://dx.doi.org/10.1016/j.physletb.2006.08.037}{Phys.\ Lett.\
  {\bf B641} (2006) 57}, \href{http://arxiv.org/abs/hep-ph/0512210}{{\tt
  arXiv:hep-ph/0512210}}\relax
\mciteBstWouldAddEndPuncttrue
\mciteSetBstMidEndSepPunct{\mcitedefaultmidpunct}
{\mcitedefaultendpunct}{\mcitedefaultseppunct}\relax
\EndOfBibitem
\bibitem{Jaeger:1402577}
A.~Jaeger {\em et~al.}, \ifthenelse{\boolean{articletitles}}{{\it Measurement
  of the track finding efficiency}, }{}
  \href{https://cdsweb.cern.ch/record/1402577?ln=en}{LHCb-PUB-2011-025}\relax
\mciteBstWouldAddEndPuncttrue
\mciteSetBstMidEndSepPunct{\mcitedefaultmidpunct}
{\mcitedefaultendpunct}{\mcitedefaultseppunct}\relax
\EndOfBibitem
\bibitem{D'Agostini}
G.~D'Agostini, \ifthenelse{\boolean{articletitles}}{{\it {A Multidimensional
  unfolding method based on Bayes' theorem}},
  }{}\href{http://dx.doi.org/10.1016/0168-9002(95)00274-X}{Nucl.\ Instrum.\
  Meth.\  {\bf A362} (1995) 487}\relax
\mciteBstWouldAddEndPuncttrue
\mciteSetBstMidEndSepPunct{\mcitedefaultmidpunct}
{\mcitedefaultendpunct}{\mcitedefaultseppunct}\relax
\EndOfBibitem
\bibitem{RooUnfold}
T.~{Adye}, \ifthenelse{\boolean{articletitles}}{{\it Unfolding algorithms and
  tests using roounfold}, }{} in {\em PHYSTAT 2011 Workshop on Statistical
  Issues Related to Discovery Claims in Search Experiments and Unfolding},
  p.~313, 2011.
\newblock \href{http://arxiv.org/abs/1105.1160}{{\tt arXiv:1105.1160}}\relax
\mciteBstWouldAddEndPuncttrue
\mciteSetBstMidEndSepPunct{\mcitedefaultmidpunct}
{\mcitedefaultendpunct}{\mcitedefaultseppunct}\relax
\EndOfBibitem
\bibitem{SVD}
A.~H\"ocker and V.~Kartvelishvili, \ifthenelse{\boolean{articletitles}}{{\it
  {SVD approach to data unfolding}},
  }{}\href{http://dx.doi.org/10.1016/0168-9002(95)01478-0}{Nucl.\ Instrum.\
  Meth.\  {\bf A372} (1996) 469},
  \href{http://arxiv.org/abs/hep-ph/9509307}{{\tt arXiv:hep-ph/9509307}}\relax
\mciteBstWouldAddEndPuncttrue
\mciteSetBstMidEndSepPunct{\mcitedefaultmidpunct}
{\mcitedefaultendpunct}{\mcitedefaultseppunct}\relax
\EndOfBibitem
\bibitem{LHCB-PAPER-2011-015}
LHCb collaboration, R.~Aaij {\em et~al.},
  \ifthenelse{\boolean{articletitles}}{{\it {Absolute luminosity measurements
  with the LHCb detector at the LHC}},
  }{}\href{http://dx.doi.org/10.1088/1748-0221/7/01/P01010}{JINST {\bf 7}
  (2012) P01010}, \href{http://arxiv.org/abs/1110.2866}{{\tt
  arXiv:1110.2866}}\relax
\mciteBstWouldAddEndPuncttrue
\mciteSetBstMidEndSepPunct{\mcitedefaultmidpunct}
{\mcitedefaultendpunct}{\mcitedefaultseppunct}\relax
\EndOfBibitem
\bibitem{HERWIG}
M.~Bahr {\em et~al.}, \ifthenelse{\boolean{articletitles}}{{\it {Herwig++
  physics and manual}},
  }{}\href{http://dx.doi.org/10.1140/epjc/s10052-008-0798-9}{Eur.\ Phys.\ J.\
  {\bf C58} (2008) 639}, \href{http://arxiv.org/abs/0803.0883}{{\tt
  arXiv:0803.0883}}\relax
\mciteBstWouldAddEndPuncttrue
\mciteSetBstMidEndSepPunct{\mcitedefaultmidpunct}
{\mcitedefaultendpunct}{\mcitedefaultseppunct}\relax
\EndOfBibitem
\bibitem{Nason:2004rx}
P.~Nason, \ifthenelse{\boolean{articletitles}}{{\it {A new method for combining
  NLO QCD with shower Monte Carlo algorithms}},
  }{}\href{http://dx.doi.org/10.1088/1126-6708/2004/11/040}{JHEP {\bf 11}
  (2004) 040}, \href{http://arxiv.org/abs/hep-ph/0409146}{{\tt
  arXiv:hep-ph/0409146}}\relax
\mciteBstWouldAddEndPuncttrue
\mciteSetBstMidEndSepPunct{\mcitedefaultmidpunct}
{\mcitedefaultendpunct}{\mcitedefaultseppunct}\relax
\EndOfBibitem
\bibitem{Frixione:2007vw}
S.~Frixione, P.~Nason, and C.~Oleari, \ifthenelse{\boolean{articletitles}}{{\it
  {Matching NLO QCD computations with parton shower simulations: the POWHEG
  method}}, }{}\href{http://dx.doi.org/10.1088/1126-6708/2007/11/070}{JHEP {\bf
  11} (2007) 070}, \href{http://arxiv.org/abs/0709.2092}{{\tt
  arXiv:0709.2092}}\relax
\mciteBstWouldAddEndPuncttrue
\mciteSetBstMidEndSepPunct{\mcitedefaultmidpunct}
{\mcitedefaultendpunct}{\mcitedefaultseppunct}\relax
\EndOfBibitem
\bibitem{Alioli:2010xd}
S.~Alioli, P.~Nason, C.~Oleari, and E.~Re,
  \ifthenelse{\boolean{articletitles}}{{\it {A general framework for
  implementing NLO calculations in shower Monte Carlo programs: the POWHEG
  BOX}}, }{}\href{http://dx.doi.org/10.1007/JHEP06(2010)043}{JHEP {\bf 06}
  (2010) 043}, \href{http://arxiv.org/abs/1002.2581}{{\tt
  arXiv:1002.2581}}\relax
\mciteBstWouldAddEndPuncttrue
\mciteSetBstMidEndSepPunct{\mcitedefaultmidpunct}
{\mcitedefaultendpunct}{\mcitedefaultseppunct}\relax
\EndOfBibitem
\bibitem{Skands:2010ak}
P.~Z. Skands, \ifthenelse{\boolean{articletitles}}{{\it {Tuning Monte Carlo
  generators: the Perugia tunes}},
  }{}\href{http://dx.doi.org/10.1103/PhysRevD.82.074018}{Phys.\ Rev.\  {\bf
  D82} (2010) 074018}, \href{http://arxiv.org/abs/1005.3457}{{\tt
  arXiv:1005.3457}}\relax
\mciteBstWouldAddEndPuncttrue
\mciteSetBstMidEndSepPunct{\mcitedefaultmidpunct}
{\mcitedefaultendpunct}{\mcitedefaultseppunct}\relax
\EndOfBibitem
\bibitem{Martin:2009iq}
A.~Martin, W.~Stirling, R.~Thorne, and G.~Watt,
  \ifthenelse{\boolean{articletitles}}{{\it {Parton distributions for the
  LHC}}, }{}\href{http://dx.doi.org/10.1140/epjc/s10052-009-1072-5}{Eur.\
  Phys.\ J.\  {\bf C63} (2009) 189}, \href{http://arxiv.org/abs/0901.0002}{{\tt
  arXiv:0901.0002}}\relax
\mciteBstWouldAddEndPuncttrue
\mciteSetBstMidEndSepPunct{\mcitedefaultmidpunct}
{\mcitedefaultendpunct}{\mcitedefaultseppunct}\relax
\EndOfBibitem
\bibitem{Lai:2010vv}
H.-L. Lai {\em et~al.}, \ifthenelse{\boolean{articletitles}}{{\it {New parton
  distributions for collider physics}},
  }{}\href{http://dx.doi.org/10.1103/PhysRevD.82.074024}{Phys.\ Rev.\  {\bf
  D82} (2010) 074024}, \href{http://arxiv.org/abs/1007.2241}{{\tt
  arXiv:1007.2241}}\relax
\mciteBstWouldAddEndPuncttrue
\mciteSetBstMidEndSepPunct{\mcitedefaultmidpunct}
{\mcitedefaultendpunct}{\mcitedefaultseppunct}\relax
\EndOfBibitem
\bibitem{Ball:2012cx}
R.~D. Ball {\em et~al.}, \ifthenelse{\boolean{articletitles}}{{\it {Parton
  distributions with LHC data}},
  }{}\href{http://dx.doi.org/10.1016/j.nuclphysb.2012.10.003}{Nucl.\ Phys.\
  {\bf B867} (2013) 244}, \href{http://arxiv.org/abs/1207.1303}{{\tt
  arXiv:1207.1303}}\relax
\mciteBstWouldAddEndPuncttrue
\mciteSetBstMidEndSepPunct{\mcitedefaultmidpunct}
{\mcitedefaultendpunct}{\mcitedefaultseppunct}\relax
\EndOfBibitem
\end{mcitethebibliography}

\end{document}